# Title: Rapid Evolution of the Photosystem II Electronic Structure during Water Splitting

**Authors:** Katherine M. Davis†, Brendan T. Sullivan, Mark Palenik ‡, Lifen Yan, Vatsal Purohit§, Gregory Robison¥, Irina Kosheleva[1], Robert W. Henning[1], Gerald T. Seidler[2], Yulia Pushkar*

**Affiliations:**

Department of Physics and Astronomy, Purdue University, West Lafayette, IN 47907, USA.

[1]Center for Advanced Radiation Sources, The University of Chicago, Chicago, IL 60637, USA.

[2]Department of Physics, University of Washington, Seattle, WA 98195, USA.

*Correspondence to: ypushkar@purdue.edu

Current addresses:

† Department of Chemistry, Princeton University, Princeton, NJ 08544, USA.

‡ Naval research laboratory, Washington, DC 20375, USA

§ Department of Biology, Purdue University, West Lafayette, IN 47907, USA.

¥ Department of Physics and Astronomy, Hanover College, Hanover, IN 47243, USA

**Abstract**: Photosynthetic water oxidation is a fundamental process that sustains the biosphere. A $Mn_4Ca$ cluster embedded in the photosystem II protein environment is responsible for the production of atmospheric oxygen. Here, time-resolved x-ray emission spectroscopy (XES) was used to observe the process of oxygen formation in real time. These experiments reveal that the oxygen evolution step, initiated by three sequential laser flashes, is accompanied by rapid (within 50 μs) changes to the Mn Kβ XES spectrum. However, no oxidation of the $Mn_4Ca$ core above the all $Mn^{IV}$ state was detected to precede O–O bond formation. A new mechanism featuring $Mn^{IV}$=O formation in the $S_3$ state is proposed to explain the spectroscopic results. This chemical formulation is consistent with the unique reactivity of the $S_3$ state and explains facilitation of the following $S_3$ to $S_0$ transition, resolving in part the kinetic limitations associated with O-O bond formation. In the proposed mechanism, O-O bond formation precedes transfer of the final ($4^{th}$) electron from the $Mn_4Ca$ cluster, in agreement with experiment.

**One Sentence Summary:** Time-resolved XES indicates that oxygen formation in Photosystem II proceeds via the reduction of the $Mn_4^{IV}$ centers, a result that has be rationalized with a molecular model.

**Main Text:**

Plants, green algae and cyanobacteria rely on the photoactive metalloprotein complex photosystem II (PS II) for its role in photosynthetic energy conversion. Using the sun's energy, PS II splits water into molecular oxygen ($O_2$), protons and electrons (*1-3*). Its catalytic activity and quantum efficiency remain unmatched by synthetic systems developed for artificial photosynthesis (*4, 5*). PS II utilizes a $Mn_4Ca$ cluster and its associated protein environment, collectively referred to as the Oxygen Evolving Complex (OEC), as the catalytic center for water splitting. In 1970, Bessel Kok *et al.* described a potential water splitting mechanism in which the OEC cycles through five states ($S_0$-$S_4$), corresponding to the oxidation states of manganese, following sequential visible light absorptions, Figure 1 (*6*). Antenna pigments from the surrounding protein matrix absorb light and funnel energy towards $P_{680}$, the chlorophyll *a* special pair responsible for charge separation. Within nanoseconds, a tyrosine residue ($Tyr_Z$) located between $P_{680}$ and the OEC is oxidized by the special pair. $Tyr_Z^\bullet$ is subsequently reduced by the OEC on a microsecond timescale. This process drives the water splitting reaction (*1*). The past forty years have yielded new insights into the structure of PS II (*3, 7-12*), as well as the nature and timescales of the S-state transitions, Figure 1 (*13-15*).

However, the critical step of O–O bond formation remains poorly characterized and thus cannot be implemented in artificial systems. The O–O bond is likely formed within microseconds during the $S_3$ to $S_0$ transient step of the Kok cycle culminating in $O_2$ evolution, but details remain unknown. A pre-eminent report by Babcock et al. yielded a rate of $Tyr_Z^\bullet$ reduction, $t_{1/2}$ ~1 ms, following three flashes and associated this rate constant with the formation of the $S_4$ state, subsequently capable of fast $O_2$ evolution (*16*). It was pointed out later that such hypotheses face a serious kinetic challenge in that the timescale for molecular oxygen release, following the formation of the $S_4$ state, is very short for the associated redox chemistry and bond formation dynamics (*2*). As an alternative hypothesis, $S_3$ state peroxide formation was proposed (*2*). However, this has not been confirmed experimentally. Given the inherent experimental complications, many computational simulations have been performed to model the O–O bond formation path (*17-20*). The majority of these imply oxidation of the OEC past $Mn_4^{IV}$ via the formation of a $Mn^V Mn_3^{IV}$ state, also presented as $Mn_4^{IV}$–$O^\bullet$ (oxyl radical). This oxidized configuration is currently associated with $S_4$ and would precede O-O bond formation (*17, 21*). Experimental proof of such an intermediate state is currently lacking and our data rule out its formation. Here, we examine the earliest dynamic in the $S_3$ to $S_0$ transition by time-resolved x-ray emission spectroscopy (TR-XES) utilizing dispersive detection to aid our understanding of this critical biological process (*22, 23*). From these results, we formulate a mechanistic hypothesis consistent with experimental data that sheds light on some of the key steps in O–O bond formation.

XES probes occupied orbitals; in particular, the Mn Kβ spectral emission lines reflect the number of unpaired 3d electrons and, thus, provide information about the oxidation and/or spin states (*24*), Figure 2 (*inset*). The exchange interaction between the 3p hole and 3d valence electrons in the final state causes multiplet splitting that results in separate $Kβ_{1,3}$ and $Kβ'$ peaks, Figure 2. This coupling is directly linked to the electronic state of Mn such that an increase in the oxidation state results in decreased splitting between the Kβ lines. This effect is apparent in the $Kβ_{1,3}$ peak position shift to lower energies with increasing oxidation. XES also allows for dispersive detection, in which the full emission spectrum is recorded during a single, intense, polychromatic x-ray pulse, Figure 2 (*22*). The temporal resolution is only limited by the time structure of the x-ray source. In our PS II experiments, we utilized multi-bunch x-ray exposures of 44 µs duration to match the microsecond kinetics of the OEC, Table 1. We previously determined that 66 µs of exposure at these conditions is undamaging to PS II (*22, 25*). Data collection was performed using a von Hamos style miniature x-ray emission spectrometer (miniXES – Figure S1) (*22, 26*), and a non-jet open-air sample delivery system (see SI and Figure S2) was used to supply fresh, unrecycled PS II for each measurement. Samples were excited given a defined number (0-3) of laser flashes (F) and probed at a time (Δt) after the final laser flash by a single x-ray pulse, Figures 2 & S3, Tables S1 & S2.

A total of six beamtimes were accomplished, with two devoted to methodology development and four to data collection, Tables 2 and S2 & S8. Time-resolved measurements following zero, one and two laser flashes were recorded at Δt=500 µs, a delay time allowing for the full reduction of Tyr$_Z^\bullet$ with limited decay of the formed S-state, Tables 2, S3, S4, S6, S8 (*13, 27*). These flash data correspond to the majority S-states $S_1$, $S_2$, and $S_3$ respectively (Table S1).

The obtained $S_1$ state spectral shape and peak position are in good agreement with previous RT PS II measurements (*22*). A comparison of 0F and 1F spectra shows a reproducible shift of the K$\beta_{1,3}$ peak to lower energies quantified by analysis of the first moment. Given the multiplet character of the spectra and the potential noise inherent for such a dilute sample, previous studies recommend the use of the statistical first moment, $\frac{\sum_j E_j \cdot I_j}{\sum_j I_j}$, surrounding the K$\beta_{1,3}$ peak (6485-6495 eV), to determine any changes to the electronic structure, Figure S8. The observed reproducible shift of the K$\beta_{1,3}$ peak to lower energies following a single laser flash is consistent with previous cryogenic XANES (*11, 28*) and XES (*29*) results for the $S_1$ to $S_2$ state transition indicating Mn-centered oxidation. In contrast, the x-ray free electron laser (xFEL)-based TR-XES analysis of the same transition in PS II did not detect this shift, likely due to poor statistics (*23*).

While the absolute shift between the first moments of 0F and 1F varies in magnitude between data sets likely due to counting statistics, Table S6 demonstrates that the values are consistently negative. Analysis of three combined data sets yields a first moment shift of -0.12 eV for the 0F → 1F transition, Table S3. Cryogenic measurements previously reported a -0.059 eV shift following spectral smoothing and background subtraction procedures (*29*). It should be emphasized that background subtraction (Figure S6) procedures can affect the first moment magnitudes, Tables S3, S6 & S7. For studies reporting small changes in the first moment values, statistical methods should be used to determine confidence in the detected changes. One-way ANOVA confirms that the first moment shift between 0F and 1F spectra detected in this study is statistically significant, Table 2. Overall, we consider our results for the 0F → 1F transition robust and in good agreement with previous characterization of the $S_1$ to $S_2$ transition in which one Mn center is oxidized from Mn$^{III}$ to Mn$^{IV}$.

For the ensuing S-state transition, $S_2 \rightarrow S_3$, previous cryogenic measurements determined that changes in the Mn K$\beta$ emission spectrum are minimal (*29*). A small (-0.02 eV) shift, on the order of our systematic error (0.02 eV, see SI for more details), was reported (*29*). Statistical analysis shows that the first moments of 1F and 2F x-ray emission spectra are indistinguishable, Table 2. Emission data representative of the $S_1$ (0F) to $S_3$ (2F) transition are presented in Figure S7. The proposed Mn oxidation to form Mn$_4^{IV}$ in the $S_3$ state (*30-33*) Figure 1, implies comparable XES shifts for the $S_1 \rightarrow S_2$ and $S_2 \rightarrow S_3$ transitions. Lack of such an effect was previously attributed to ligand-centered oxidation (*29*). However, we know that the OEC undergoes a major structural change during the $S_2 \rightarrow S_3$ transition from both EXAFS (*28*) and femtosecond (fs) x-ray crystallography (*12*). If such a structural change significantly alters the hybridization of the Mn 3d and ligand orbitals, some neutralization of the Mn-centered oxidative shift could be proposed to explain these results.

To probe samples enriched with $S_0$, XES spectra were collected at a 40 ms delay following a third laser flash, ($3F_{40ms}$). The first moment of $3F_{40ms}$ is consistently shifted to higher energies (Tables S3-S6) and statistically distinct from 1F and 2F first moments, Table 2, supporting the expected reduction of Mn. Overall, the RT TR-XES results are in good agreement with Mn Kβ emission data for previously cryogenically trapped states (*29*). Having validated the experimental technique, we investigated the elusive transient $S_3$ to $S_0$ process also initiated by three laser flashes (3F).

Figure 3 depicts the earliest evolution of the OEC electronic structure following three laser flashes and the associated first moment shifts of Mn $K\beta_{1,3}$ measured at Δt ~50 μs ($3F_{50\mu s}$) and Δt ~200 μs ($3F_{200\mu s}$). The trend of increasing first moment is robust and statistically significant, Table 2. Furthermore, the observed increase in the first moment cannot be explained by mixing of states only, see the SI for details regarding laser excitation. This result suggests that during the $S_3$ to $S_0$ transition the OEC undergoes a significant transformation at short timescales. Note that changes between 2F and $3F_{200\mu s}$ are greater than between 2F and $3F_{500\mu s}$. We currently attribute this to limited statistics or to transient oxidation of the OEC by the electron arriving from Tyr$_Z^\bullet$. Transient oxidation would likely reverse the reductive trend in first moments to one of increasing value.

Contrary to the density functional theory (DFT)-based modeling predictions of O–O bond formation (*17-20*), we were unable to observe oxidation of the OEC, which would be associated with lower values of the first moment, at any observation time point following the third flash and measured over multiple beamtimes. This lack of evidence for oxidation past 2F (majority state $S_3$) was consistently observed (Figure 3 & Tables S3–S7) thereby excluding formation of a $Mn^V Mn_3^{IV}$ state kinetic intermediate. Current DFT models (*17-20*) also do not resolve the kinetic challenge identified by Prof. Renger (*2*). UV-Vis difference spectra show that the electron leaves Tyr$_Z^\bullet$ quite slowly, ~1 ms after the third flash. Given that this time constant is comparable to the rate of $O_2$ evolution, only a very short ~50 μs time window remains for all bond formation dynamics and product/substrate exchange to occur.

While our results cannot be explained by existing DFT mechanistic models, they *are* in agreement with the only other published TR study probing the electronic structure evolution of the Mn centers via x-ray absorption spectroscopy (XAS) (*13, 34*). TR-XAS detected no oxidation and instead, suggests the gradual (milliseconds) reduction of the OEC initiated 250 μs into the $S_3$ to $S_0$ transition. In contrast to the XAS study, where only two energy points along the Mn K-edge were analyzed, we collect full spectra representing the complete electronic structure of the OEC at Δt ~50 μs and Δt ~200 μs. In addition, XAS and XES represent different electronic transitions. It is therefore possible that some early spectral changes may have previously escaped detection. A more recent XES study performed at the Linac Coherent Light Source indicates no changes to the Mn Kβ spectra 250 μs after the third laser excitation (*35*). As with the unobserved $S_1$ to $S_2$ shift at the xFEL, we attribute this discrepancy to the differences in experimental conditions that will likely be clarified in the future.

Figure 4 introduces a mechanism based on the fs-XRD structure of PS II (*36*) and verification of structural models via EXAFS (*37*) that is consistent with recent experiments. DFT calculations in Gaussian09 (*38*) were used to perform geometry optimizations. We hypothesize that the anomalous O5 in the XRD structure corresponds to an OH bridge between $Mn_{4A}$ and $Mn_{3B}$ (also proposed by Prof Shen (*36*)). However, we assign it as a structural element of the $S_0$ state similar to the $S_0$ structure presented by Krewald et al. (*39*). While not critical for the main conclusions of this manuscript, this hypothesis implies that the crystals studied might have been produced partially or completely in the $S_0$ state. Subsequent oxidation of $S_0$ and deprotonation of the OH bridge give rise to an $S_1$ state structure that agrees well with recent room temperature EXAFS (*37*). To ensure equivalency between energy inputs for each S-state transition in our model, we describe S-state transitions as proton-coupled electron transfer (PCET) events, Figure 4. A summary of the structural changes inherent to our mechanistic model is in Table S9. In agreement with experiment, the $S_1$ to $S_2$ transition results in minimal changes. The fundamental hypothesis in our proposed mechanism is the presence of a $Mn^{IV}=O$ fragment in the $S_3$ state, located on the last Mn center to be oxidized, $Mn_{1D}$. Our model suggests that this center cannot oxidize in lower S-state transitions as it lacks a ligand capable of PCET. The high activation energy measured for the $S_2$ to $S_3$ transition (*40*) is in good agreement with the need for substrate binding to $Mn_{1D}$ and substantial structural changes to create a reactive $Mn^{IV}=O$ fragment. The model also predicts significant shortening of one Mn-Ca vector previously observed experimentally (*28*), Table S9. The presence of $Mn^{IV}=O$ agrees well with the distinct chemical reactivity of the $S_3$ state in comparison to other S-states (*41*), provides rationale for the suppression of oxidative shifts in XES by changes in orbital hybridization, and decreases the "kinetic challenge" (*2*) associated with the $S_3$ to $S_0$ transition. The presence of a substrate oxygen as a $Mn^{IV}=O-Ca$ bridge is also in agreement with the order of magnitude increase in substrate exchange rate upon Ca to Sr substitution (*42*). Single point energies were obtained for a broken symmetry calculation of $S_3$. Diagonalizing the Heisenberg-Dirac-Van Vleck Hamiltonian yielded an $S_3$ ground state spin of 3, in agreement with the most recent EPR data (*33*), as well as indicating radicaloid character of the oxygen in the $Mn^{IV}=O-Ca$ fragment due to its high spin density of $\rho=0.4$. The potential for O–O bond formation was subsequently analyzed for $Mn_{1D}^{IV}=O + H_2O-Ca$ and $Mn_{4A}-O-Mn_{3B} + Mn_{1D}^{IV}=O$, as these correspond to shortest O–O distances in the $S_3$ state model, Figure 4. Both were found to be larger than the driving force available from the Tyr–OH=Tyr–O$^{\bullet}$ + e$^-$ + H$^+$ couple. (Table S10?) However, in the case of O–O bond formation between $Mn_{1D}^{IV}=O$ and $H_2O-Ca$, the increase in energy can be compensated by binding of a new water molecule to Ca according to the equation: $[Mn_{1D}^{IV}=O] + [H_2O-Ca] + H_2O = Mn_{1D}-OOH + [H_2O-Ca]$. While this scheme is formally equivalent to the use of $H_2O$ from the protein environment and preserves Ca–$H_2O$ coordination, it seems less likely due to the absence of free water in close proximity to $[Mn_{1D}^{IV}=O]$. Interestingly, the energy rise was similar with and without the abstraction of an electron, Table S10.

$$[Mn_{1D}^{IV}=O] + [H_2O-Ca] + H_2O = Mn_{1D}-OOH + [H_2O-Ca]+1e^-+1H^+$$
$$[Mn_{1D}^{IV}=O] + [H_2O-Ca] + H_2O = Mn_{1D}-OOH + [H_2O-Ca]+1H^+$$

DFT calculations show an increase in required energy input for each following S-state transition, even when they are conducted similarly (as PCET), Figure 4. In manmade catalysts, the consecutive oxidations of the catalytic center required for O–O bond formation also necessitate the application of progressively increasing oxidation potentials. For instance, ruthenium-based water oxidation catalysts undergo several redox transitions such as $Ru^{II}/Ru^{III}$, $Ru^{III}/Ru^{IV}$, and $Ru^{IV}$/the catalytic state, occurring at progressively increasing redox potentials

(*43, 44*). PCET can decrease the differences in consecutive oxidation steps but cannot completely eliminate it. The OEC works under a fixed potential provided by the Tyr–OH=Tyr–O$^\bullet$ + e$^-$ + H$^+$ couple. Thus, any variation in redox potential between the S-state transitions will result in energy losses unless excess energy from low oxidation state (such as $S_0/S_1$ and $S_1/S_2$) transitions can be efficiently stored in the form of an altered protein conformation (mechanical energy) or as a re-distribution of charges (electrical energy). Our simplistic analysis of reaction energetics can be refined in the future by including the larger protein environment using molecular dynamics simulations in conjunction with TR-fs-XRD of S-state transitions.

To overcome the kinetic challenge, we propose that the O-O formation step takes place in the $S_3$ to $S_0$ transition prior to the reduction of TyrY$_Z^\bullet$. Rapid evolution of the OEC during the $S_3$ to $S_0$ transition has long been a primary target of PS II research. Based on UV-Vis difference spectroscopy (*15, 45*) and TR infrared spectroscopy (*14*) a deprotonation event has been proposed to occur early (0-300 μs) during this transition. Our results do not explicitly exclude a deprotonation event, but necessitate significant changes to the electronic structure of the 3d Mn frontier orbitals to explain the observed spectroscopic effect. The results presented here can be better rationalized if the formation of the (TyrY$_Z^\bullet$)$S_3$ state, occurring on the order of 100ns (*14*), triggers a sequence of events resulting in significant redox or structural changes to the OEC, such as the formation of the O–O bond. The most recent isotope exchange studies show that substrate exchange stops early (0-300 μs) in the transition (*42*) hinting at such a possibility.

The paramount question is whether the O–O bond can be formed in the Mn$_4$Ca cluster prior to electron transfer to TyrY$_Z^\bullet$. This possibility would require significant energy input such as from a strong base or proton-pump to drive deprotonation, mechanical relaxation of the protein resulting in O–O bond formation, and/or stabilization of the peroxide product. Stored energy from $S_0$ to $S_3$ conversion can be as high as ~2.5 eV if no energy dissipation occurs, which is sufficient for O–O bond formation by proton abstraction, Figure 4, S9. In this case, the final step of O$_2$ evolution would correspond to the oxidation of the Mn$_4$Ca(OOH) cluster by TyrY$_Z^\bullet$. A particularly interesting feature of the proposed (TyrY$_Z^\bullet$)$S_3$OOH state structure is the hydrogen bond between the peroxide and the Mn4A–O–Mn3B bridge that is protonated in the $S_0$ state. Thus, deprotonation of the -OOH fragment can take place under oxidation to result in (TyrY$_Z$)$S_4$(O$_2$)$^{\bullet-}$, from which expulsion of O$_2$ follows directly resulting in $S_0$ and providing a logical conclusion of the cycle.

In summary, to analyze the evolution of the photosystem II electronic structure, we observed intermediate states of photosynthetic O$_2$ production via microsecond resolution time-resolved X-ray emission spectroscopy at a synchrotron source. Consistent with the obtained full spectra is a mechanism involving the presence of Mn$^{IV}$=O in the $S_3$ state, and O–O bond formation in the $S_3$ to $S_0$ transition prior to TyrY$_Z^\bullet$ reduction. This mechanism resolves the previously highlighted kinetic problems and opens to investigation the possibility of energy storage in the surrounding protein matrix. Mn$^{IV}$=O–Ca activated fragments mimicking the OEC have already been demonstrated in synthetic systems (*46*) and the use of such fragments to drive O–O bond formation will likely be discovered in the future.

**Acknowledgments:** The spectrometer and TR-XES methodology development was supported by the DOE, Office of Basic Energy Sciences DE-FG02-12ER16340 (Y.P.). Measurements of Photosystem II were supported by NSF, CHE-1350909 (Y.P.) and the NSF Graduate Research Fellowship under Grant No. DGE0833366 (K.D.). Research at the University of Washington is supported by the DOE, Office of Basic Energy Sciences DE-SC0002194. PNC/XSD facilities at the Advanced Photon Source and research at these facilities are supported by the U.S. Department of Energy, Basic Energy Sciences, a Major Resources Support grant from NSERC, the University of Washington, Simon Fraser University, and the Advanced Photon Source. Use of the Advanced Photon Source, an Office of Science User Facility operated for the U.S. Department of Energy (DOE) Office of Science by Argonne National Laboratory, was supported by the U.S. DOE under Contract No. DE-AC02-06CH11357. Use of the BioCARS Sector 14 was also supported by grants from the National Center for Research Resources (5P41RR007707) and the National Institute of General Medical Sciences (8P41GM103543) from the National Institutes of Health. The time-resolved setup at BioCARS was funded in part through a collaboration with Philip Anfinrud (NIH/ NIDDK). We thank Prof. L. Slipchenko from Purdue University for providing computational resources and helpful discussion.

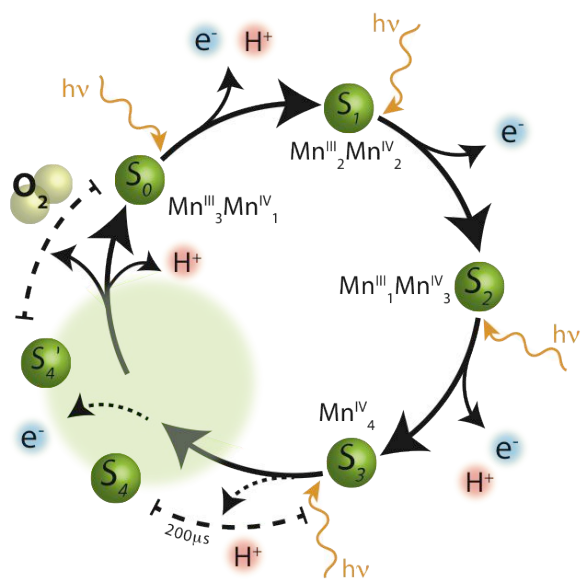

**Fig. 1.** A current model of the Kok cycle depicting incident visible light photons and electron/proton release (*13*). The dashed region is based on previous analysis of the $S_3$ to $S_0$ transition in which the $S_4$ state was proposed (*13, 34*).

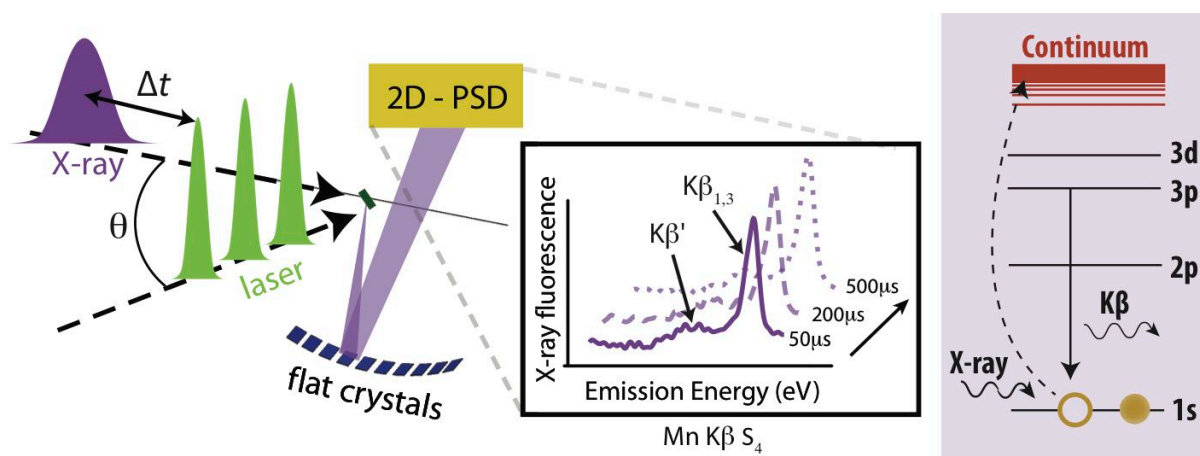

**Fig. 2.** Schematic of the experimental setup. Nanosecond laser pulses (1, 2 or 3) are used to advance the Kok cycle in the protein, Table S2. The pump/probe delay time, $\Delta t$, measured from the final laser flash to the center of the x-ray pulse, is set dependent on the desired S-state, Figure S3 and Table S2. X-ray fluorescence from the sample is reflected by 10 flat analyzer crystals onto a 2D-position sensitive detector. K$\beta$ emission spectra are extracted to form snapshots of the electronic structure in time. Smoothed emission spectra are presented for the $S_3$ to $S_0$ transition.

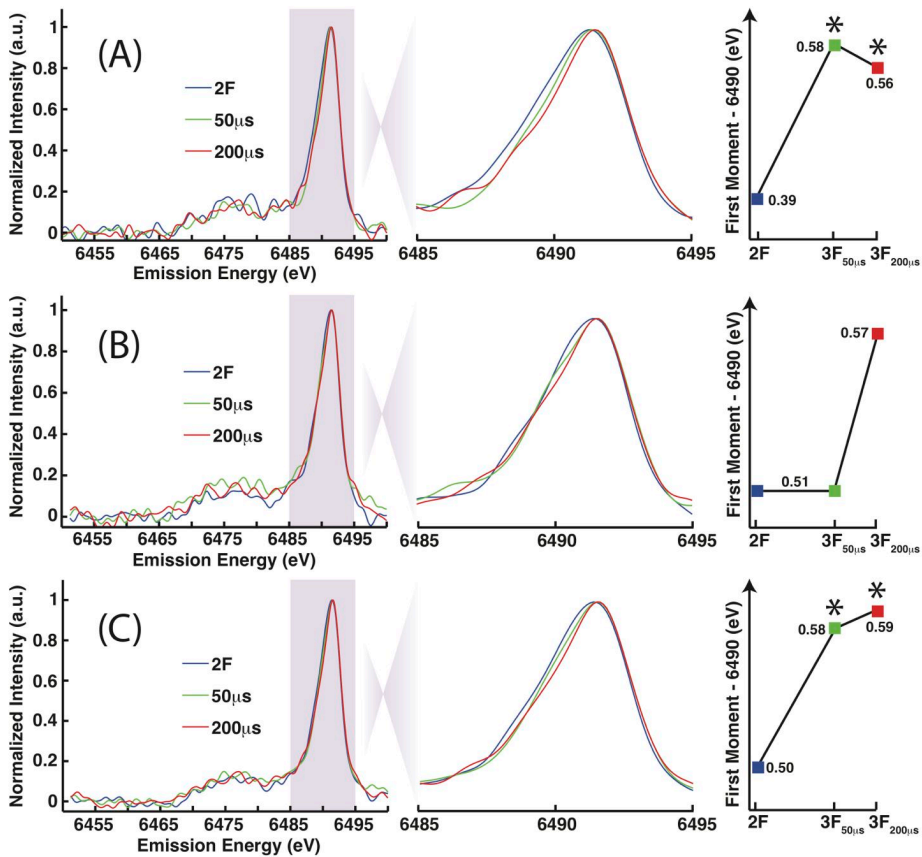

**Fig. 3.** Fully processed (background subtracted and smoothed) 2F, $3F_{50\mu s}$, and $3F_{200\mu s}$ Kβ emission spectra for data sets (A) 3 (including the additional statistics, Table S7), (B) 4 and (C) their sum respectively. The region (6.485-6.495 keV) over which the first moment was calculated is highlighted and a magnified inset presented. For each set of spectra the trend in first moments is shown. Note that variations between data sets are due to counting statistics that cause noise on the edges of the selected region. Those moments statistically different from 2F data are indicated with an asterisk. See Tables S6 and S7 for the first moment magnitudes.

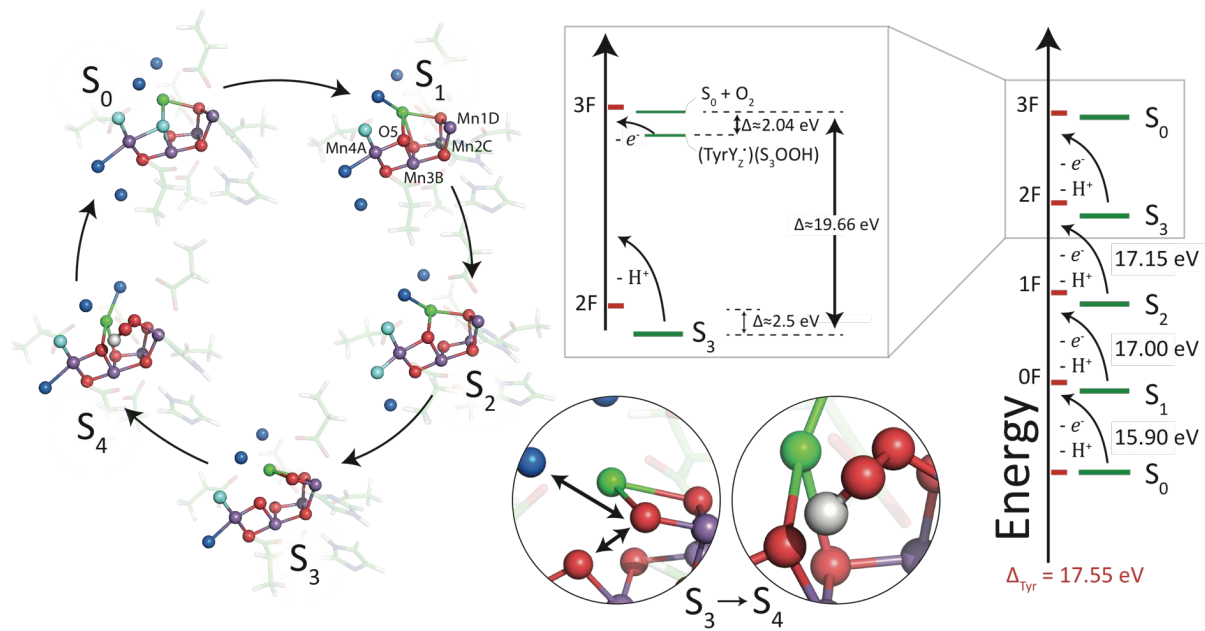

**Fig. 4.** A mechanism of water oxidation has been developed to explain the spectroscopic results. Left: The orientation of molecular models are aligned for clarity. Purple, red and green atoms are Mn, O, and Ca respectively. Water molecules are shown in dark blue, hydroxides are light blue. Amino acids shown are those used in the DFT calculations. Bonds are shown for distances <2.5 Å. Magnifications of $S_3$ and $S_4$ show conversion of the $Mn^{IV}$=O-Ca reactive fragment to the peroxo species hydrogen bonding to the $Mn_{4A}$–O–$Mn_{3B}$ bridge. Note that the $S_3$ inset is rotated slightly from the common alignment to allow for easier visualization of the $Mn^{IV}$=O-Ca fragment. Arrows indicate the shortest distances most likely for O-O bond formation: $Mn_{1D}^{IV}$=O + $H_2O$–Ca and $Mn_{4A}$–O–$Mn_{3B}$ + $Mn_{1D}^{IV}$=O.

Right: On the energy diagram red ticks represent the energy change ($\Delta_{Tyr}$) due to $TyrY_Z\cdot$ reduction in the PCET process: (Tyr–OH=Tyr–O· + $e^-$ + $H^+$). Green ticks show the changes in energy due to S-state transitions. The $S_3$ to $S_0$ transition is shown in greater detail including the energy levels of proposed intermediates.

**Table 1**. Experimental characteristics of pulsed x-ray source

| Characteristics | BioCARS |
|---|---|
| Excitation Energy | Peak energy 7.85 keV, FWHM ~500 eV |
| X-ray Spot Size | ~45 x 100 μm$^2$ |
| Pulse Length | 44 μs |
| Photon Flux | 3 x 10$^{11}$ photons/pulse |
| Dose Delivered per Pulse | ~7 x 10$^7$ photons/μm$^2$ |
| Repetition Rate | ~20 Hz |

**Table 2.** ANOVA values for each S-state calculated from raw (unprocessed) data.

| Majority S-State | | $S_0$ | $S_1$ | $S_2$ | $S_3$ | $S_{4a}$ | $S_{4b}$ | $S'_4$ |
|---|---|---|---|---|---|---|---|---|
| | Flash | $3F_{40ms}$ | 0F | 1F | 2F | $3F_{50\mu s}$ | $3F_{200\mu s}$ | $3F_{500\mu s}$ |
| $S_0$ (117) | $3F_{40ms}$ | 1 | 0.25 | 0.05 | 0.03 | 0.69 | 0.41 | 0.10 |
| $S_1$ (78) | 0F | | 1 | <0.01 | <0.01 | 0.46 | 0.78 | 0.02 |
| $S_2$ (82) | 1F | | | 1 | 0.98 | 0.03 | 0.01 | 0.98 |
| $S_3$ (97) | 2F | | | | 1 | 0.01 | <0.01 | 0.96 |
| $S_{4a}$ (76) | $3F_{50\mu s}$ | | | | | 1 | 0.68 | 0.08 |
| $S_{4b}$ (100) | $3F_{200\mu s}$ | | | | | | 1 | 0.04 |
| $S'_4$ (84) | $3F_{500\mu s}$ | | | | | | | 1 |

*p-values are based on the first moments over the range 6.485 – 6.495 keV, and are calculated for all data, i.e. including every beamtime. The number of "samples" (i.e. threads) is shown in parentheses for each state. See Table S8 for additional comparisons between the states based on the number of X-ray pulses per state per beamtime and Figure S8 for a dotplot of all first moments used for p-value calculations.

**Supplementary Materials:**

Materials and Methods, optimized DFT coordinates

Figures S1-S9

Tables S1-S10

References (*47-55*)

# Supplementary Materials

## Materials and Methods

**Photosystem II Preparation:**

PSII-enriched thylakoid-membrane particles were prepared from supermarket spinach (*47, 48*). Prior to use, samples were stored at -80°C in a buffer using sucrose as a cryo-protectant: 0.4 M sucrose, 5 mM $CaCl_2$, 5 mM $MgCl_2$, and 15 mM NaCl, 50 mM MES, pH 6.0. The oxygen evolution activity of PSII was measured by a Clark-type electrode in a Hansatech oxygraph. The activity of the preparation was 300 μmol $O_2$/(mg Chl • hr) or greater under constant saturating illumination at 25°C utilizing 0.3 mM 2,6-dichloro-1,4-benzoquinone (DCBQ) as an artificial electron acceptor. The Chl a:b ratio was derived from the optical absorbance of chlorophyll extracted with 80% acetone: 20% water solution. This was measured with a Cary300 Bio UV-visible spectrophotometer. For all samples, this ratio was ~2.5:1 which indicates high enrichment of membrane particles with PSII. To ensure the high quality of the samples, low temperature X-band EPR spectra were recorded for the $S_1$ and $S_2$ states. $S_2$ state samples were obtained by illuminating $S_1$ state samples with 120 W Halogen lamp for 30 minutes while in a cold bath of ethanol and dry ice to maintain a sample temperature of 195 K, after which they were immediately frozen in liquid nitrogen.

For TR-XES, the PSII samples were prepared on-site immediately prior to the measurements as follows. After thawing the stock pellet (~30mg Chl/mL) stored on dry ice for travel, the samples were diluted 1:1 with a 30% v/v glycerol in a buffer of 5 mM $CaCl_2$, 5 mM $MgCl_2$, and 15 mM NaCl, 50 mM MES, pH 6.5 and homogenized gently with a paintbrush. A solution of 50 mM PPBQ in DMSO was added to yield a final PPBQ concentration of 500 μM (*28, 49*). Clean mixed fiber thread of radius ~175 μm was tested to have no Mn and remain intact under laser and x-ray illuminations. The sample was painted onto these threads with a soft brush. Painting ensures best sample absorption onto the threads. Threads were wound onto small plastic spools and servo motors were used to slowly wind and unwind the threads between the plastic spools at a constant speed during painting. During measurements, the spool with freshly painted thread was placed on an axis above a reservoir of ice and covered with aluminum foil. Measurements were performed as soon as possible (<30 min from application of sample to data collection). All sample preparation, handling and storage environments were completely dark, save for dim green LEDs when unavoidable, to prevent state transitions prior to measurement. In addition, samples were maintained at a constant temperature of approximately 4°C during preparation.

**Beamline Description:**

TR emission spectra were collected at sector 14-ID B, BioCARS (*50*), of the Advanced Photon Source, Argonne National Lab with an electron energy of 7.0 GeV and an average current of 100 mA. To avoid second order reflections and gallium fluorescence inherent to the spectrometer crystals, the undulator gap of U27 was set to 11.7 mm during the emission experiments yielding a ~500 eV FWHM undulator spectrum centered around 7.85 keV. In addition, the water-cooled KB mirrors for focusing were translated to their Si stripes and set to an angle of 3.8 mrad, thereby cutting the incident energy range sharply above the energy of the first harmonic. The vertical mirror was defocused to yield a final projection size of ~45 x 100 μm$^2$ (V x H) given the sample surface angled 45° to the incident beam, and the incident flux was monitored downstream of both mirrors via a photodiode. X-ray pulses of ~44 μs were produced using the high heat load chopper (*50*) and delivered at frequency 10.3 or 20.6 Hz, see Table S2 for experimental parameters.

Monochromatic beam was produced using a Si(111) channel-cut monochromator inserted into the beam path and calibrated via the $KMnO_4$ pre-edge at 6.5433 keV. Monochromatized beam is necessary for the energy calibration of the position sensitive detector. The smallest possible undulator gap (10.5 mm), corresponding to a peak energy of 6.85 keV, was used such that the tails of the undulator spectrum reach 6.445 keV, the lowest collection energy of the GaP spectrometer. To improve the energy resolution of x-rays reflected by the monochromator, the beam incident on the channel-cut was collimated by decreasing the radius of curvature of the vertical mirror. The scattering signal used for calibration was collected with 1 kHz pulse frequency to increase the integrated flux.

**XES Spectrometer:**

For this study, we utilized a short working distance (SWD) miniature x-ray emission spectrometer (miniXES) in which multiple flat dispersive GaP 110 analyzers reflect x-ray fluorescence onto a Pilatus 100k (Dectris) 2-dimensional position sensitive detector (2D-PSD) Figure S1 (*22, 25*). To decrease attenuation, the internal spectrometer space is purged with He during use. The spectrometer crystals are arranged in a von Hamos geometry and designed for a collection range of ~50 eV about the Kβ main lines, with each crystal reflecting the complete Mn Kβ spectrum (6445-6510 eV) onto the PSD, Figure S1. The width of monochromatic elastic scatter measured by the spectrometer is a convolution of the monochromator spectral resolution ($\Delta E/E = 1.4 \times 10^{-4}$) and the broadening introduced by the spectrometer itself, determined to be ~0.3 eV. Note that this resolution is distinct from the minimum energy shifts resolvable by the setup; our maximum obtained resolution is discussed in **Data Processing / Averaging**.

Prior to measurements, the spectrometer was aligned horizontally and vertically via eyelets designed for this purpose. The thread position with respect to the spectrometer in the beam direction (see more in **Sample Positioning**) was optimized prior to PS II data collection using a sample with intense Mn Kβ fluorescent signal, typically $MnCl_2$ solution painted on the threads. Multiple small (~100 μm) sample translations along the beam direction were implemented until all 10 crystals were fully visible on the Pilatus face. The thread position in the beam direction was fixed using the focus of a stationary, high magnification camera. A lens extension with a focal depth of ~50μm was used to ensure consistent spatial positioning.

**Sample Positioning:**

To provide the least amount of stray scatter into the spectrometer aperture, the samples were tilted vertically 45° towards the incident beam, Figure S2(A). Kinematic mounts were used to attach the sample delivery system, Figure S2(A), to a motorized X,Y,Z stage. These mounts provide a simple, highly reproducible method for exchanging samples with minimal shifts of thread position relative to the spectrometer. Immediately following the sample mounting procedure, the thread was aligned with the x-ray beam via a computerized transmission scan perpendicular to the beam (inboard/outboard – $\hat{x}$ in Figure S1) using the downstream diode (Thorlabs PDA36A). To prevent the destruction of the diode, a reduction in beam intensity was necessary in addition to a protective layer of attenuation foil on the diode. Once the center of the thread was confirmed horizontally, the thread was brought into the focal point of the camera using the vertical and along-the-beam motors ($\hat{y}$ and $\hat{z}$ respectively, in Figure S1), with the inboard/outboard ($\hat{x}$) position held constant. Typically, a few small steps of 50-100 μm were required.

The sample delivery system, Figure S2(A), includes plastic goal posts protruding from an aluminum plate and angled 45° to the beam. Stainless steel syringe needles of radius 203 μm are securely fitted through holes in these posts to deliver sample reproducibly at the beam position. The needles are positioned far outside of the spectrometer's acceptance volume to prevent contamination from ancillary Mn. A stepper motor is mounted on the upstream side of the sample assembly, and an aluminum spool of starting radius 9.55 mm or 11.2 mm (data set #1 only, see Table S2 for an experimental summary) is attached to the motor to pull and collect thread as it passes through the interaction region. See Figure S2(A). In general, sample positions were not changed except upon mounting and aligning in small increments as determined by transmission scan and by the camera system, see more in **Establishing the Interaction Region and Timing Schemes**. Note that the sample coated spool was covered completely with aluminum foil during the positioning time to prevent any unintentional x-ray or light exposure that could alter the protein state prematurely. Once the sample position was finalized, this covering was partially removed to allow free translation of the thread, and the collection spool was rotated repeatedly until the previously enclosed fresh sample reached the beam position.

**Spectrometer Calibration:**

An *in situ* calibration of the detector pixels is achieved by measuring the positions of the elastic scattering peaks while scanning the monochromator through the x-ray emission energy range (6445-6510 eV). The calibration methodology has been described in detail elsewhere (*26, 51, 52*). Note that the line width of the elastic peaks is the convolution of the monochromator and spectrometer energy resolutions. It is important to recognize that changes in sample position relative to the spectrometer alter the detector calibration. We thus developed a system to reproducibly position the sample (tolerance <50 μm), and calibrated the spectrometer daily following the completion of measurements on a given sample (typical sample consisted of 6-8 m of painted thread). In addition to this, $MnCl_2$ solution emission spectra were collected after each sample on the same thread without any sample repositioning. These spectra were used as an internal reference.

**Establishing the Interaction Region and Timing Schemes:**

*Pump/probe Alignment and Camera Setup:*

In addition to the criticality of reproducible sample position in the miniXES system, exact alignment between laser and x-ray is crucial for accurate TR data. Light excitation was provided by a Spectra-Physics Empower laser (527 nm, pulse length ~200 ns) placed at an angle θ~5° to the incident x-ray beam. Additional optics allowed for a FWHM spot size of ~220 μm, at a laser fluence of ~11 mJ/mm$^2$. The laser spot size was selected to be about twice larger than x-ray spot size to ensure best alignment. The x-ray beam was visualized on a phosphor (SPI P47 coated scintillator) positioned at the previously established sample position relative to the spectrometer and the laser was moved to an overlapping position.

To facilitate fine adjustments to the sample alignment and monitor the interaction region during data collection, a Prosilica camera directed at the sample was mounted vertically on a small X,Y,Z stage. To improve our precision, a lens extension with a focal depth of ~50 μm was used. Once in place, the focal region of the camera was aligned to a test thread known to be in alignment with the spectrometer setup. The thread was then removed and replaced by the phosphor mounted at the same position, see Figure S2. It is critical that the phosphor surface appears in focus on the camera. X-ray/laser alignment was checked periodically using the phosphor. No deviations were found during 4 days of beamtime. Once these alignments are complete, fine adjustments of the sample/spectrometer alignment for optimal emission signal are made by small spectrometer translations only. From that point the sample position is 'fixed' at the focal region of the camera, the spectrometer position is finalized, and data collection starts.

*X-ray/Laser Synchronization:*

For verification of timing, the phosphor was replaced by a small Si photodiode aligned at the beam position (in the focus of the camera). Synchronization between the pump (laser) and probe (x-ray) pulses was performed by recording the respective diode signals on an oscilloscope. A field programmable gate array (FPGA) was used to maintain the timing between the laser and x-ray pulses. The programmable delay between laser and x-rays pulses was adjusted to overlap in time the peaks of each pulse that was subsequently defined as time zero. Given the broad 44 µs FWHM X-ray and ~200 ns FWHM laser pulses, it is more exact to synchronize via peak position instead of rising edges. It follows that spectra recorded at a set delay include fluorescence from ±20 µs, see Figure S3.

**Photosystem II advancement in the Kok cycle:**

In this work we do not deconvolute the XES spectra but instead analyze changes in the first moments as a function of the number of laser flashes and time delay to the probe pulse. For a prediction of S-state composition, we refer to Han et al. who recently established the maximal experimentally achievable S-state composition after each laser flash at room temperature for PS II membrane preparations from spinach, Table S1 (*49*). To ensure the most efficient S-state conversion, we emulate previous experimental laser settings. At a laser fluence of ~11 mJ/mm$^2$, assuming the threads are fully saturated with sample, the ratio of incident photons to PS II centers is ~560. This matches the only other XES experiment by Messinger et al. in which advancement to ~65% of the $S_0$ state was determined by EPR (*29*). Given we are able to probe the created states long before any decay occurs, unlike the manual freezing in Messinger et al., we set this value as a lower bound on our $3F_{40ms}$ data. This convolution of states will be similar for each of the $S_4$ time delays, given they are pumped with the same number of laser flashes. It is clear, therefore, that any reductive effects we observe are in reality more significant than they appear. It should also be emphasized that convolution with $S_{2,3}$ data is not likely to lessen any oxidative effect more than ~30%.

**Data Collection:**

Data collection scans were synchronized in the following manner – for parameters α, β and γ see Table S2: (i) the stepper motor translates the sample α microns through the beam position transporting fresh material into the beam; (ii) the specified number of laser flashes are delivered to the fresh sample; (iii) subsequently, the ~44 µs x-ray pulse arrives at a delay, β; (iv) immediately following the closure of the shutter, the cycle repeats. Note that both laser and x-ray pulse at the same frequency γ. During the repetition time, the Pilatus continuously collects a single image. Sample threads were mounted in ~5 to ~10 m lengths and monitored every ~0.5 m resulting in multiple Pilatus images per generated state. These images were directly summed during data processing for improved statistics.

## Data processing / averaging

*Averaging and a Systematic Error Estimation*

To evaluate the effects of adding up data obtained on different (and thus re-positioned relative to spectrometer) threads, we first analyzed the strong $MnCl_2$ spectra. Highly concentrated $MnCl_2$ solution (~1 M) was painted onto each thread at the conclusion of PS II measurements prior to sample exchange, and a single PSD image was collected. First moments were calculated around the range 6490-6495 eV (as the $MnCl_2$ peak is higher in energy than PS II) and a standard deviation of 0.02 eV was found. To verify that summing threads would not affect the resolution, we also checked with data that had been split randomly into two groups and observed shifts between 0.005 – 0.02 eV. Consistent with both these methods, we consider 0.02 eV to be our maximal resolution for PSII XES collection. Note also that averaging of multiple threads broadens the data slightly due to minor variations in thread position. These variations are random and therefore can be assumed to be even about the $K\beta_{1,3}$ peak with limited effect on the first moments. Given the sensitivity to sample positioning, each thread was used to collect all S-states/intermediates and only one calibration file was used to process the data reducing the error due to the calibration files themselves. However, calibrations were still taken daily to ensure large shifts in the monochromator calibration or other major effects did not occur. To completely reduce any effects of position or monochromator calibration drift on our S-state shifts, we collected all spectra in approximately equally intensities (equal number of detector exposures) for every thread. It follows that any changes in the absolute position of the sample do not affect the relative shifts between the states.

*Processing PS II spectra*

A total of six beamtimes were accomplished, with two devoted to methodology development and four to data collection. For every beamtime (data sets 1 – 4, Table S2), each TR XES spectrum is a direct summation of the Pilatus exposures for all individual threads using a single calibration. Differing data sets are also summed for greater statistics. Results from different beamtimes were merged using the strong $MnCl_2$ signal as an internal energy-calibration reference. Similar to PS II, all $MnCl_2$ spectra in the data set are summed and calibrated with one calibration file.

Statistical first moments were analyzed for PS II data processed in three different ways: completely unprocessed, with background subtraction only, and after smoothing and background removal, Tables S6 & S7. No significant differences were found in the first moment trends between these methods; errors introduced by processing do not alter unprocessed data trends that are significant to within a 95% confidence interval, see Table 2.

Examples of data for the $S_3$ to $S_0$ transition are shown in Figure S4 and first moments of the raw data are in Table S3-S7. Due to high levels of background, and relatively low statistics, consistent background removal was necessary to reliably compute first moment values for comparison with other published works both past and future. A quartic polynomial was fit to the spectral region excluding the Kβ peaks, Figure S6. The computed polynomial was then subtracted from the spectrum. Note that removal of a linear background was found to create significant variability in the first moments and is thus insufficient. Data were then normalized to the $K\beta_{1,3}$ peak maximum for the purpose of comparison. For both raw and background subtracted data some variability in the first moment trends between beamtimes persists due to noise in the spectral range. To combat this problem, smoothing techniques were tested.

*Wavelet Transforms*

The wavelet transform method was chosen to smooth non-physical rapid fluctuations in the data while maintaining the integrity of the spectral shape.
The wavelet transformation of our spectrum $f(E)$ is (*53*):

$$w(a,E) = C_\psi^{-1/2} a^{-1/2} \int_{-\infty}^{\infty} \psi^* \left( \frac{E' - E}{a} \right) f(E') dE' \tag{1}$$

where $C_\psi$ is a normalization constant. The base wavelet function, $\psi_{a,b}(E) = \psi\left(\frac{E-b}{a}\right)$, is dilated and translated to allow for decomposition of the spectrum. Dilation is described by $a$; $b$ characterizes translation. For our purposes, we use the Mexican Hat wavelet function, see equation 2 (*53*).

$$\psi(E) = (1 - E^2) e^{-E^2/2} \tag{2}$$

Wavelet functions are calculated for the range of $a$ 0.01 to 30, over the interval 6453.7 eV to 6513 eV. The energy-frequency plane of $w(a,E)$ allows for visualization of the most appropriate cutoff for the dilation parameter $a$ used in the spectral reconstruction. Figure S5 provides an example spectrogram. Note that the intensity in z is determined by the $w(a,E)$ value (*53*).

Our chosen parameters leave the K$\beta_{1,3}$ peak adequately smoothed but without significant broadening while the K$\beta$' peak maintains some noisy character. It is possible to completely eradicate noise in K$\beta$' peak, but the process results in significant broadening and symmetrizing of the K$\beta_{1,3}$ peak, an effect we want to avoid as it can skew the first moment calculations. Figure S6 compares the wavelet reconstruction of $3F_{500\mu s}$ state data to the 'raw' spectrum.

**Error Determination:**

For some experiments, it is common to present the standard error of the mean (SEM) from the second moment, essentially an evaluation of the error on the first moment (our measure of spectral position, mathematically the weighted arithmetic mean). The SEM can be written as follows:

$$\text{SEM} = \sqrt{\frac{\mu_2}{N}} \text{ and } \mu_2 = \frac{\sum_i (E_i - \bar{E})^2 I_i}{\sum_i I_i} \tag{3}$$

$\mu_2$ is the weighted second moment calculated from the intensities, $I_i$, and emission energies, $E_i$, at every point, as well as the weighted first moment, $\bar{E}$. $N$ is the number of observations. Given that the SEM is highly dependent on the sampling/binning of data into $N$ groups, we prefer to avoid this error analysis method. Additionally, SEM results can be misleading in that overlapping error bars do not necessarily exclude statistically significant differences.

*Analysis of Variance*

**AN**alysis **O**f **VA**riance (ANOVA) between experimental groups was used to determine the confidence level for the observed first moment shifts between states. One-way ANOVA, calculated from the first moments of raw unprocessed normalized data, yields the p-values in Table 2. These values represent the probability of the null hypothesis, i.e. that a difference in the first moment between the two states listed is a random statistical variation. In this study we take the 95% confidence interval (1.96σ) to be significant.

The first moment trends for raw data match well with those for smoothed, background-subtracted spectra (i.e. 'processed'), Table S3 within our confidence levels, Table 2. In other words, for those states in which we likely reject the null hypothesis, smoothing and background subtraction do not alter the trends in the changes to the first moments. We present first moment shifts for data collected on the same threads, Tables S3, S5, & S7. However, ANOVA analysis between the same states (excluding 2F) from different beamtimes shows no statistical variation, p≥0.19, Tables 2 & S4. The only outlier 2F spectrum from data set 3 has a dip on the low energy side of the Kβ$_{1,3}$ peak between 6.485 – 6.488 keV. This noise causes a significant variation in first moments between data sets and skews the mean of summed data to lower energies. Unfortunately, this spectral noise is not easy to remove without completely excluding the 2F data from that beamtime, after which statistical invariance is re-established. However, in the full data set, Table S4, inclusion versus exclusion of beamtime #3 2F data only changes the first moment by 0.01eV, again, below our systematic error. To preserve statistics we include these data.

**DFT Calculations:**

Undamaged coordinates from the fs-XRD structure were used as the initial geometry (*36*). Geometry optimizations were performed in Gaussian09 (*38*) on a 94 atom model of S$_3$ including the Mn core and a portion of the surrounding backbone using the bp86 (*54*) functional and def2 tzvp basis set (*55*). Eight single point calculations representing the possible relative spin orientations of the Mn centers with the spin lying along a single axis were used to build the Heisenberg-Dirac-Van Vleck (HDVV) Hamiltonian, where the total energy is given as

$$H = E_0 - \sum_{ij} J_{ij} \hat{S}_i \cdot \hat{S}_j \quad (4)$$

Here, $E_0$ is a constant background energy, $\hat{S}_j$ and $\hat{S}_j$ are the spin operators of the individual Mn centers with indices i and j, and $J_{ij}$ is the coupling constant between them. The values of $J_{ij}$ and $E_0$ were determined by minimizing the means squared error in energy between the HDVV Hamiltonian and the energies obtained through DFT. The S$_3$ ground state spin and energy were obtained from the lowest eigenvalue of the fitted HDVV Hamiltonian.

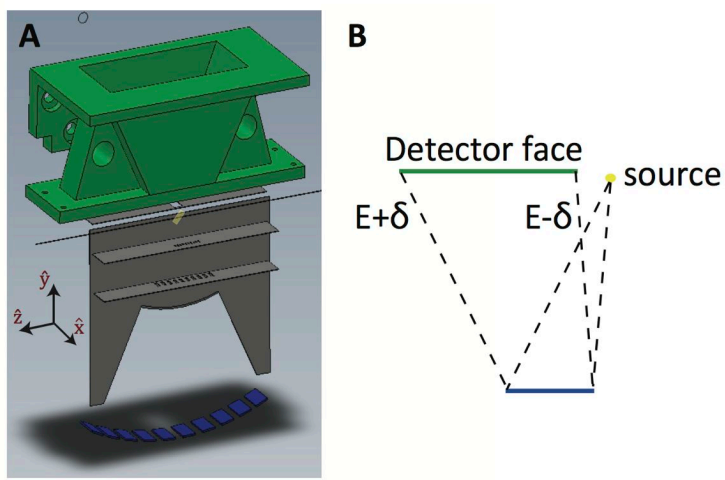

**Fig. S1.** A) 3D model of the miniXES spectrometer in the von Hamos configuration using the GaP 440 Bragg reflections. Multiple crystals are used for increased solid angle collection and are positioned for optimal use of the detector. The yellow square represents the sample plane. (B) Schematic explanation of the dispersive direction on the detector face due to the Bragg reflections.

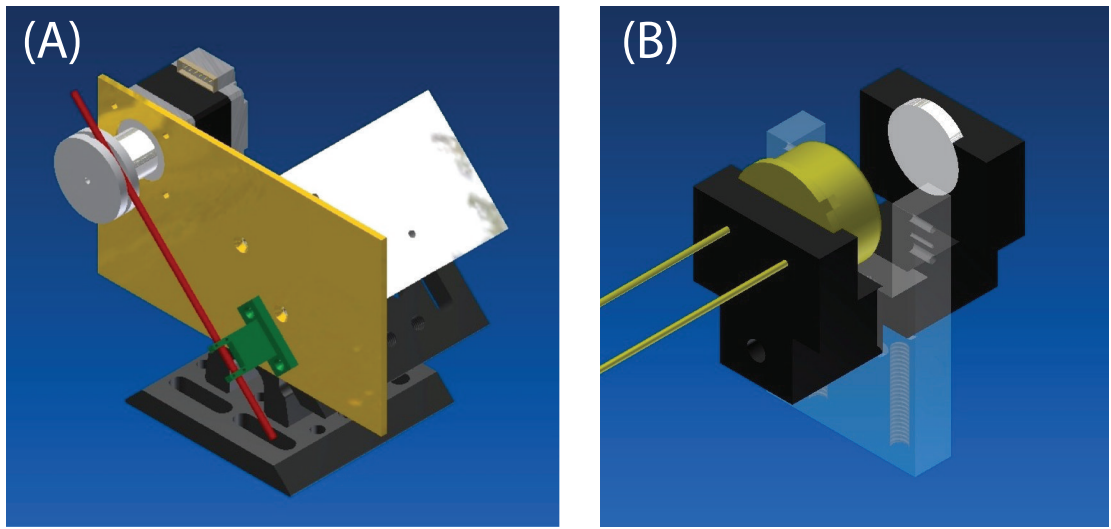

**Fig. S2.** Sample delivery and sample positioning system. (A) Plastic (green) goalposts are mounted at 45° to the incident beam. Sample coated thread (red) passes through two narrow steel needles embedded as guides in the plastic piece (green). A computer controlled step motor rotates a spool in defined intervals to translate the sample and collect the thread after it passes the interaction region between the goalposts. The kinematic mount is below the white plate to allow for reproducible sample exchange. (B) The plastic piece (green) in (A) can be exchanged to allow for alignment equipment to be placed at the exact sample position. Here is a visualization of both the small timing photodiode (gold) and the alignment phosphor and how they attach to the apparatus.

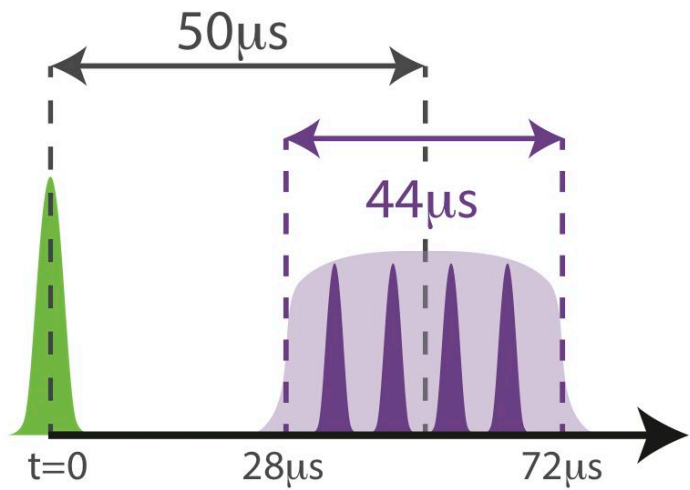

**Fig. S3.** Timing schematic showing the example delay time 50 μs. Laser (green)/X-ray (purple) synchronization was done via the pulse centers. The number of x-ray bunches (dark purple) per 44 μs pulse is variable depending on storage ring parameters during each beamtime. Data sets 1-3 had approximately 4000, data set 4 only ~300.

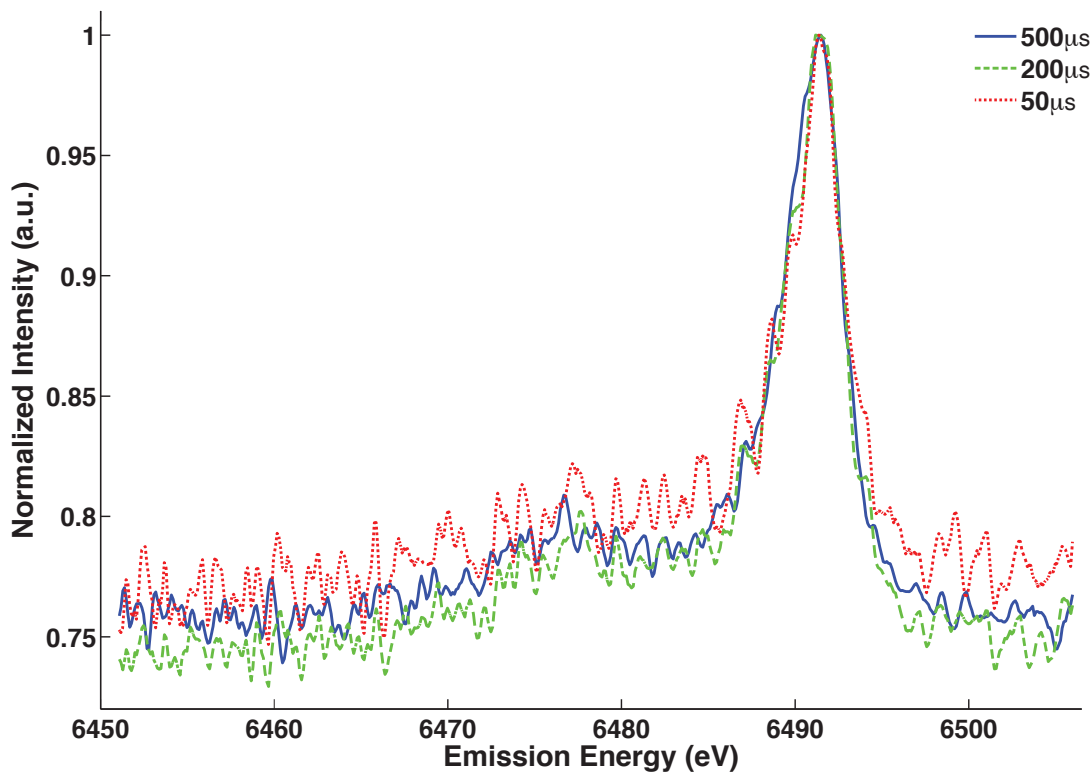

**Fig. S4.** Mn Kβ XES for OEC during the S$_3$ → S$_0$ transition at delay times of 500 μs, 200 μs, and 50 μs from the third laser flash. Data from different beamtimes are aligned using the MnCl$_2$ standard. S/N reflects the amount of data collected per state and improves with the merging of several data sets. The 500 μs spectrum is a sum of data sets 1 – 3; the 200 μs spectrum includes data from data sets 2 and 3; and the 50 μs data set is from data set 3 only. See Tables S2 & S8 for details. Background levels from stray scatter entering the detector are significant, making up approximately three quarters of the total signal. It is clear, however, that the signal to background remains essentially static with little beamtime-to-beamtime variation.

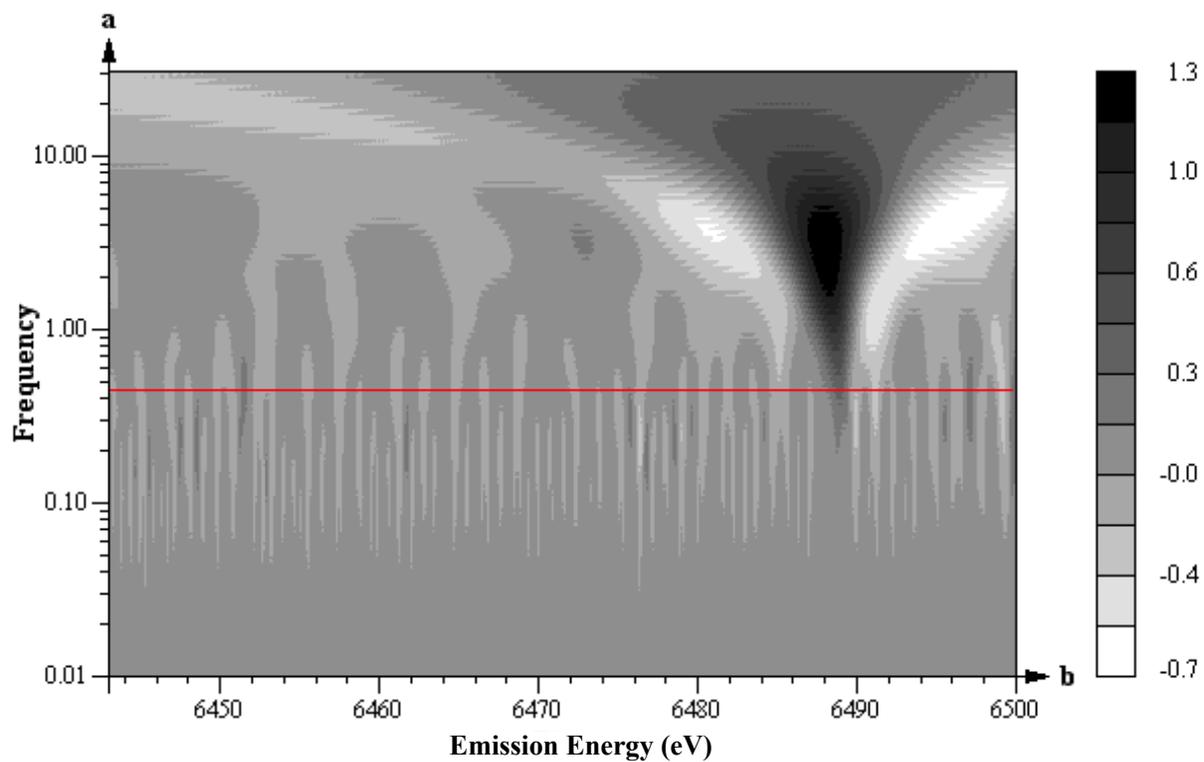

**Fig. S5.** Example spectrogram for the 500 μs raw data from data set 2. The red line is our chosen frequency cutoff, below which we begin to lose spectral features or experience spectral broadening due to the wavelet reconstruction. The grey-scale color is based on the $w(a,E)$ value at each point.

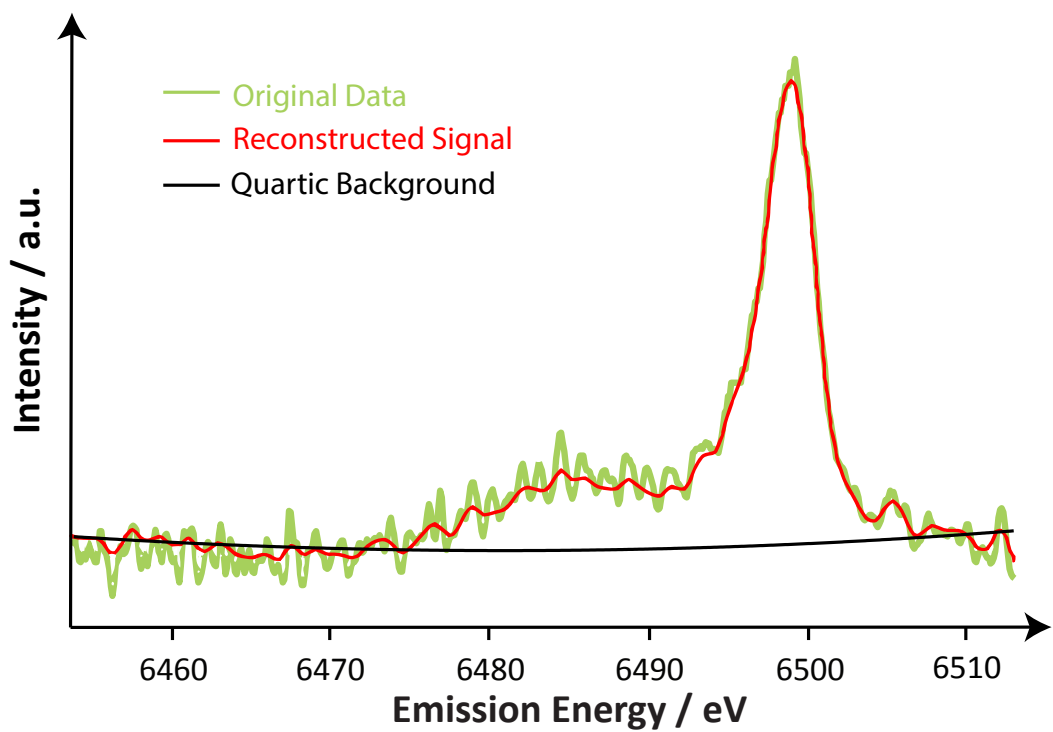

**Fig. S6.** Example of unprocessed $3F_{500\mu s}$ data from data set 2 overlaid with its wavelet-reconstructed spectrum. The quartic background (black line) shown was calculated for the reconstructed signal.

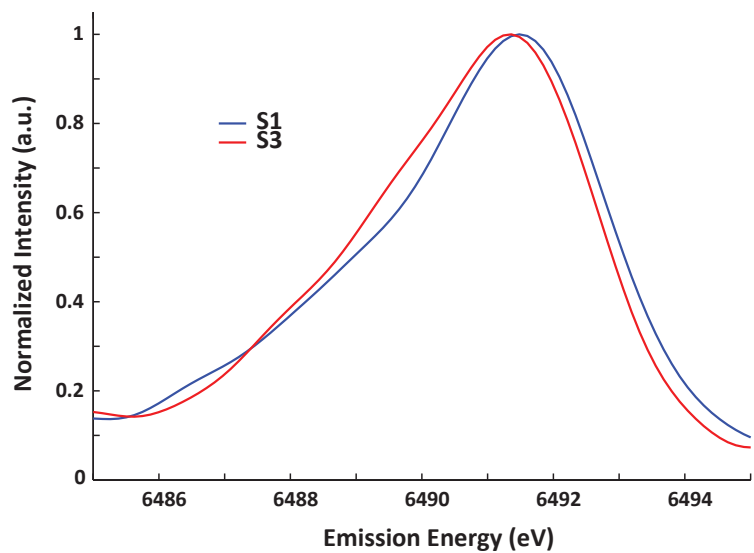

**Fig. S7.** Fully processed (background subtracted and smoothed) 0F (majority $S_1$) versus 2F (majority $S_3$) Kβ emission spectra for data sets 2 and 3 collected on the same threads.

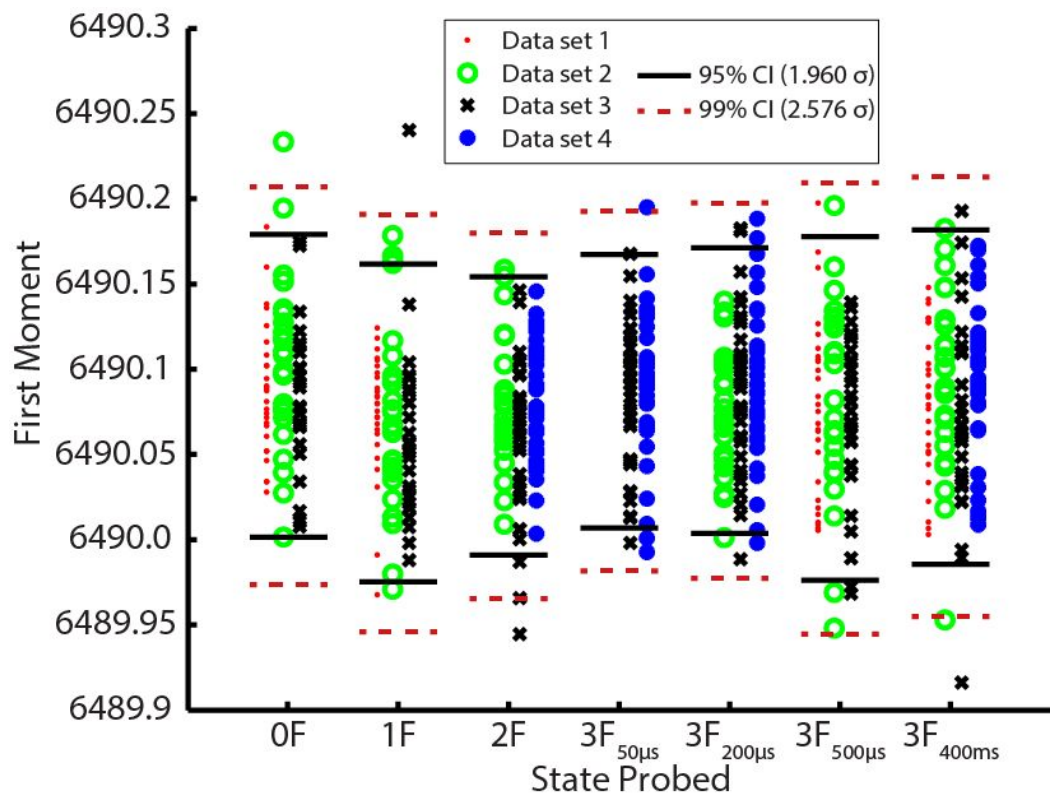

**Fig S8.** Dot plot of all calculated first moments from raw data. Each dot represents a calculated first moment from a thread collected during the beamtime corresponding to its shape in the legend. The bars represent the 95% (solid) and 99% (dashed) confidence intervals.

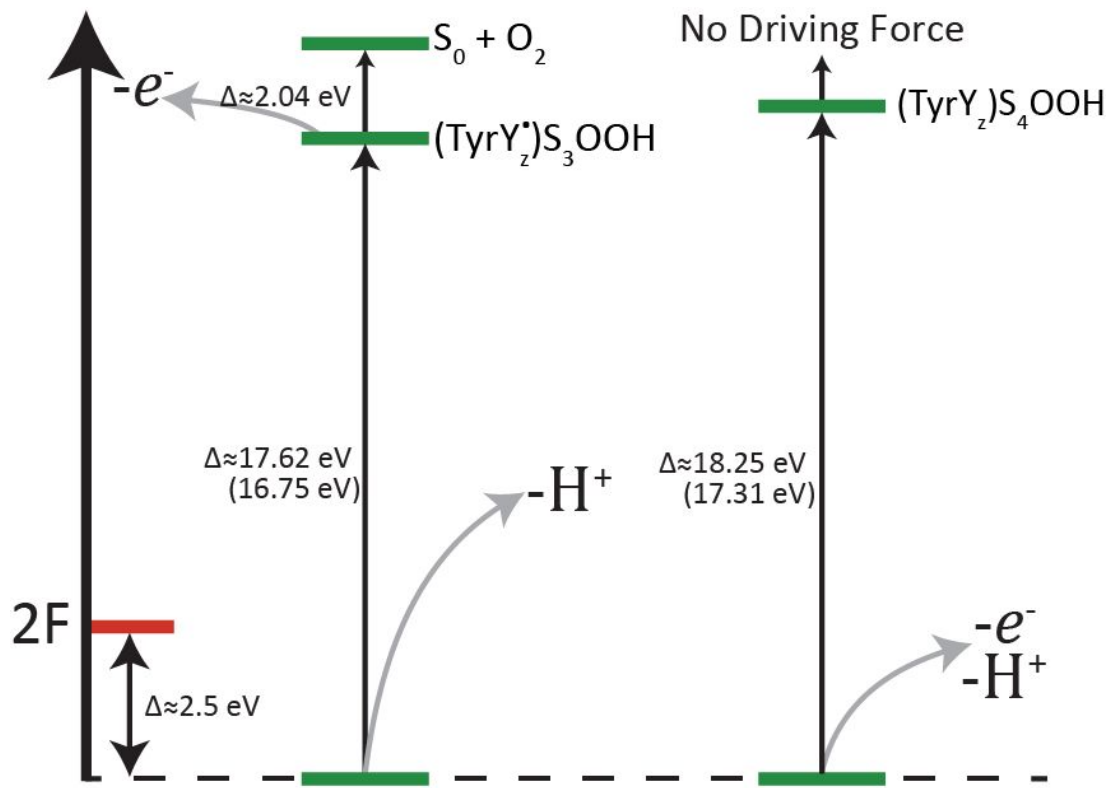

**Fig S9.** Visualization of the key energetic steps in the water oxidation mechanism from Table S10. Energies in parentheses represent the energies obtained from broken symmetry DFT calculations. This figure is an addition to Figure 4 showing more details.

**Table S1.** Maximal experimentally achievable S-State composition at room temperature.

| Flash No. | $S_1$ | $S_2$ | $S_3$ | $S_0$ |
|---|---|---|---|---|
| 0 | 100 | | | |
| 1 | 10 | 90 | | |
| 2 | | 23 ± 1 | 76 | |
| 3 | | 5 | 22 ± 1 | 74 |

*As determined by Han et al. (*49*).

**Table S2. S-state dependent scan parameters**

| State | No. of Laser Pulses | $\alpha$* (μm) | $\beta$** (μs) | $\gamma$*** (Hz) | Data Sets |
|---|---|---|---|---|---|
| $S_0$ | 3 | 600 / 450 | 40,000 | 10 | 1 – 4 |
| $S_1$ | 0 | 300 / 300 | n/a | 20 | 1 – 3 |
| $S_2$ | 1 | 600 / 450 | 500 | 20 | 1 – 3 |
| $S_3$ | 2 | / 450 | 500 | 20 | 2 – 4 |
| $S_{4a}$ | 3 | /450 | 50 | 20 | 3 & 4 |
| $S_{4b}$ | 3 | / 450 | 200 | 20 | 2 – 4 |
| $S_4'$ | 3 | 600 / 450 | 500 | 20 | 1 – 3 |

* Translation in μm (spool radius: 11.2 / 9.55 mm, for data sets 1 / 2,3,4)
** Delay Time
*** Laser and x-ray Frequency

**Table S3.** First moments of unprocessed versus processed data.

| Flash | Majority S-State | Unprocessed Data $1^{st}$ Moments – 6490 eV for beamtimes: | | Processed Data $1^{st}$ Moments – 6490 eV for beamtimes: | |
|---|---|---|---|---|---|
| | | 1 – 3 | 2 & 3 | 1 – 3 | 2 & 3 |
| $3F_{40ms}$ | $S_0$ | 0.085 | 0.087 | 0.497 | 0.474 |
| 0F | $S_1$ | 0.093 | 0.095 | 0.541 | 0.518 |
| 1F | $S_2$ | 0.069 | 0.066 | 0.418 | 0.359 |
| 2F | $S_3$ | n/a | 0.068 | n/a | 0.401 |
| $3F_{200\mu s}$ | $S_{4b}$ | n/a | 0.090 | n/a | 0.526 |
| $3F_{500\mu s}$ | $S'_4$ | 0.078 | 0.081 | 0.485 | 0.448 |
| Beamtimes | Majority S-state Trends | | | | |
| 1 – 3 | $S_2 \to S'_4 \to S_0 \to S_1$ | | | $S_2 \to S'_4 \to S_0 \to S_1$ | |
| 2 & 3 | $S_2 \to S_3 \to S'_4 \to S_0 \to S_{4b} \to S_1$ | | | $S_2 \to S_3 \to S'_4 \to S_0 \to S_1 \to S_{4b}$ | |
| Max data | $S_3 \to S_2 \to S'_4 \to S_0 \to S_{4b} \to S_1$ | | | $S_3 \to S_2 \to S'_4 \to S_0 \to S_{4b} \to S_1$ | |

*First moments are calculated over the range 6.485 – 6.495 keV. Absolute magnitude variations between processed and unprocessed data occur primarily due to background removal. Only states taken on the same threads are compared, excluding the row labeled 'max data' which assumes the statistical invariance between beamtimes to compare the maximum collected statistics for each state.

**Table S4.** First moments of *unprocessed* data from all beamtimes.

| Flash | Majority S-State | 1st Moments – 6490 eV | |
|---|---|---|---|
| $3F_{40ms}$ | $S_0$ | 0.085 | |
| 0F | $S_1$ | 0.088 | |
| 1F | $S_2$ | 0.066 | |
| 2F | $S_3$ | 0.070 | 0.071 |
| $3F_{50\mu s}$ | $S_{4a}$ | 0.086 | 0.088 |
| $3F_{200\mu s}$ | $S_{4b}$ | 0.088 | 0.088 |
| $3F_{500\mu s}$ | $S'_4$ | 0.0750 | |

*First moments are calculated over the range 6.485 – 6.495 keV. All beamtimes are included for comparison due to the statistical invariance between the states. Additional statistics were collected for $S_3$, $S_{4n}$ and $S_4$ for data set 3, see Table S8. The columns are split to reflect the adjusted first moments including these statistics.

**Table S5.** First moments of *unprocessed* 50μs delay data.

| Flash | Majority S-State | 1st Moments – 6490 eV for beamtimes: | |
|---|---|---|---|
| | | 2, 3 & 4 | 3 & 4 |
| $3F_{40ms}$ | $S_0$ | 0.086 | 0.084 |
| 2F | $S_3$ | 0.069 | 0.067 |
| $3F_{50\mu s}$ | $S_{4n}$ | n/a | 0.085 |
| $3F_{200\mu s}$ | $S_4$ | 0.088 | 0.092 |

*First moments are calculated over the range 6.485 – 6.495 keV. Only states taken on the same threads are compared in each column.

**Table S6.** First Moments – 6490 eV for individual data sets.

| Majority S-State | Flash | Raw | | | | Raw – background | | | | Processed | | | |
|---|---|---|---|---|---|---|---|---|---|---|---|---|---|
| | | \multicolumn{12}{c}{Data Sets} |
| | | 1 | 2 | 3 | 4 | 1 | 2 | 3 | 4 | 1 | 2 | 3 | 4 |
| $S_0$ | $3F_{40ms}$ | 0.083 | 0.092 | 0.072 | 0.093 | 0.55 | 0.49 | 0.42 | 0.63 | 0.53 | 0.53 | 0.44 | 0.57 |
| $S_1$ | 0 | 0.088 | 0.104 | 0.074 | n/a | 0.57 | 0.49 | 0.58 | n/a | 0.55 | 0.53 | 0.53 | n/a |
| $S_2$ | 1 | 0.075 | 0.067 | 0.058 | n/a | 0.51 | 0.34 | 0.34 | n/a | 0.49 | 0.35 | 0.38 | n/a |
| $S_3$ | 2 | n/a | 0.076 | 0.051 / 0.060 | 0.082 | n/a | 0.43 | 0.34 / 0.41 | 0.59 | n/a | 0.42 | 0.28 / 0.39 | 0.51 |
| $S_{4a}$ | $3F_{50\mu s}$ | n/a | n/a | 0.080 / 0.088 | 0.090 | n/a | n/a | 0.57 / 0.61 | 0.58 | n/a | n/a | 0.57 / 0.58 | 0.51 |
| $S_{4b}$ | $3F_{200\mu s}$ | n/a | 0.079 | 0.090 / 0.086 | 0.093 | n/a | 0.43 | 0.68 / 0.58 | 0.64 | n/a | 0.44 | 0.66 / 0.54 | 0.57 |
| $S'_4$ | $3F_{500\mu s}$ | 0.073 | 0.084 | 0.070 | n/a | 0.53 | 0.44 | 0.41 | n/a | 0.50 | 0.46 | 0.41 | n/a |

*First moments are calculated over the range 6.485 – 6.495 keV. Only states taken on the same threads are compared in each column. Additional statistics were collected for $S_3$, $S_{4n}$ and $S_4$ for data set 3. The columns are split to reflect the adjusted first moments.

**Table S7.** First moments -6490 eV for averaged data sets 3 and 4.

| Majority S-State | Flash | Raw | Raw-background | Processed |
|---|---|---|---|---|
| $S_3$ | 2 | 0.069 | 0.51 | 0.50 |
| $S_{4a}$ | $3F_{50\mu s}$ | 0.088 | 0.60 | 0.58 |
| $S_{4b}$ | $3F_{200\mu s}$ | 0.091 | 0.62 | 0.59 |

*First moments are calculated over the range 6.485 – 6.495 keV. Only states taken on the same threads are compared in each column. These are calculated including the additional statistics collected for data set 3.

**Table S8.** Approximate number of x-ray pulses per state per beamtime.

| Majority S-State | Flash | Data Sets | | | | | | Total | |
|---|---|---|---|---|---|---|---|---|---|
| | | 1 | 2 | 3 | | 4 | | | |
| $S_0$ | $3F_{40ms}$ | 43,250 | 60,000 | 55,333 | | 107,222 | | 265,805 | |
| $S_1$ | 0 | 62,267 | 72,800 | 60,233 | | n/a | | 195,300 | |
| $S_2$ | 1 | 42,083 | 61,556 | 55,333 | | n/a | | 158,972 | |
| $S_3$ | 2 | n/a | 62,000 | 57,178 | 48,889 | 107,222 | 226,400 | 275,289 | |
| $S_{4a}$ | $3F_{50\mu s}$ | n/a | n/a | 57,178 | 48,889 | 111,667 | 168,845 | 217,734 | |
| $S_{4b}$ | $3F_{200\mu s}$ | n/a | 60,889 | 57,178 | 48,889 | 110,556 | 228,623 | 277,512 | |
| $S'_4$ | $3F_{500\mu s}$ | 43,750 | 60,889 | 57,400 | | n/a | | 162,039 | |

\* Additional statistics were collected for $S_3$, $S_{4n}$ and $S_4$ for data set 3. The columns are split to reflect the additional x-ray pulses for these states. Unless specified, figures and calculations did not reflect these data to preserve the comparison between threads.

**Table S9.** Interatomic distances (Å) based on the optimized coordinates for each S-state of the proposed model.

|  | $S_0$ | $S_1$ | $S_2$ | $S_3$ | $S_4OOH + H_2O$ |
|---|---|---|---|---|---|
| **Mn1-Mn2** | 2.84 | 2.74 | 2.76 | 2.81 | 2.79 |
| **Mn1-Mn3** | 3.50 | 3.31 | 3.39 | 3.57 | 3.68 |
| **Mn1-Mn4** | 5.43 | 4.87 | 5.16 | 5.34 | 5.56 |
| **Mn2-Mn3** | 2.73 | 2.79 | 2.80 | 2.80 | 2.86 |
| **Mn2-Mn4** | 5.24 | 5.10 | 5.25 | 5.14 | 5.19 |
| **Mn3-Mn4** | 2.93 | 2.72 | 2.78 | 2.71 | 2.74 |
| **Mn1-Ca** | 3.58 | 3.50 | 3.48 | 3.14 | 3.58 |
| **Mn2-Ca** | 3.27 | 3.31 | 3.32 | 3.37 | 3.32 |
| **Mn3-Ca** | 3.45 | 3.40 | 3.35 | 3.45 | 3.34 |
| **Mn4-Ca** | 3.88 | 3.61 | 3.80 | 3.87 | 3.67 |

**Table S10.** DFT energies and spin states

| | E (Hartree) | ΔE (eV) | Spin State | Broken Symmetry E (Hartree) | Broken symmetry ΔE (eV) | Broken symmetry spin |
|---|---|---|---|---|---|---|
| $S_0$ | -8023.4311 | n/a | 1/2 | | | |
| $S_1$ | -8022.8449 | 15.95 | 1 | | | |
| $S_2$ | -8022.2197 | 17.01 | 1/2 | | | |
| $S_3$ | -8021.5906 | 17.12 | 3 | -8021.5878 | | 3 |
| $S_3OOH$ | -8097.4059 | | 1 | -8097.4351 | | 1 |
| $S_3 + H_2O = S_3OOH/H_2O + H^+$ | | 17.62 | | | 16.75 | |
| $S_3 = S_4OOH + H^+ + e^-$ | | 18.76 | | | | |
| $S_4(Mn_{3B}\text{-}OO\text{-}Mn_{1D})$ | -8020.9200 | | 1/2 | -8020.9514 | | 1/2 |
| $S_3 = S_4(Mn_{3B}\text{-}OO\text{-}Mn_{1D}) + H^+ + e^-$ | | 18.25 | | | 17.31 | |
| $S_4OOH + H_2O$ | -8097.4031 | | 1/2 | -8097.4206 | | 1/2 |
| $S_3 + H_2O = S_4OOH/H_2O + H^+ + e^-$ | | 17.70 | | | | |
| $S_3OOH/H_2O = S_4OOH/H_2O + e^-$ | | 0.08 | | | 0.39 | |

\* Broken symmetry calculations only available for $S_3$ and $S_4$ states. Unless it is specified, ΔE is the total change in energy from the previous state according to equation $S_{n+1} = S_n + H^+ + e^-$, see also Fig. 4 for scheme of Kok cycle. Whenever a proton and electron were removed, ΔE includes their energies. Such a presentation avoids estimating separate proton energies, which might have large errors.

# Optimized Cartesian Coordinates

## S₀

| | | | | | | | |
|---|---|---|---|---|---|---|---|
| C | -6.63321 | 1.87587 | 1.01510 | O | -0.81702 | 0.58051 | -1.10075 |
| C | -5.86435 | 0.56014 | 0.90321 | O | -2.06168 | 4.61179 | 1.86295 |
| C | -4.49165 | 0.62932 | 0.22296 | O | -3.29780 | -3.62029 | -1.17834 |
| O | -3.84687 | -0.43384 | 0.10380 | O | -2.92807 | -0.70277 | -2.77683 |
| O | -4.11284 | 1.79073 | -0.18391 | O | -2.97370 | 1.98652 | -2.67907 |
| C | -0.84672 | -4.82847 | -4.55234 | O | -3.41675 | 4.43536 | -0.51320 |
| C | 0.32852 | -3.90080 | -4.25068 | H | -5.68826 | 0.11960 | 1.89841 |
| C | 0.13495 | -2.97397 | -3.04387 | H | -6.44881 | -0.19404 | 0.35077 |
| O | -0.93416 | -3.02102 | -2.38902 | H | -6.82841 | 2.30943 | 0.02375 |
| O | 1.12588 | -2.18747 | -2.80521 | H | -1.04236 | -5.50798 | -3.71026 |
| C | 4.13044 | 1.14442 | -3.62258 | H | -1.76681 | -4.25444 | -4.73037 |
| N | 4.92460 | 1.05067 | -2.49280 | H | 1.25365 | -4.47329 | -4.07014 |
| C | 3.04035 | 0.34667 | -3.37052 | H | 0.55144 | -3.25307 | -5.11515 |
| C | 4.30444 | 0.23243 | -1.60546 | H | 2.19179 | 0.11495 | -4.00526 |
| N | 3.15783 | -0.20941 | -2.10918 | H | 4.71211 | -0.01402 | -0.62928 |
| C | 5.97422 | 1.41024 | 1.56162 | H | 4.80357 | -0.33576 | 2.40190 |
| N | 5.56022 | 2.52946 | 0.86325 | H | 2.83474 | 0.83000 | 1.21552 |
| C | 4.89235 | 0.60538 | 1.86969 | H | 3.55425 | 3.07477 | 0.26803 |
| C | 4.23878 | 2.39067 | 0.76184 | H | 5.17295 | -3.49244 | 1.36990 |
| N | 3.78931 | 1.25234 | 1.35992 | H | 3.89296 | -4.71073 | 1.54706 |
| C | 4.18780 | -3.68410 | 1.81550 | H | -4.00160 | -4.50730 | 2.72276 |
| C | 3.15024 | -2.72654 | 1.24167 | H | -3.43853 | -5.77028 | 3.85057 |
| O | 2.25630 | -2.31764 | 2.05254 | H | -2.86839 | -5.74930 | 2.16437 |
| O | 3.25416 | -2.42910 | 0.01175 | H | -1.08540 | -4.82498 | 3.72816 |
| C | -1.96284 | -4.22591 | 3.43695 | H | 2.29604 | 1.62516 | 4.86056 |
| C | -1.50124 | -3.22785 | 2.36778 | H | 0.97903 | 2.82088 | 4.62824 |
| O | -2.15796 | -3.09582 | 1.30882 | H | -0.25242 | 0.81081 | -1.85890 |
| C | -3.13451 | -5.11268 | 3.02099 | H | -1.73511 | 3.69740 | 1.53870 |
| O | -0.43568 | -2.58321 | 2.70127 | H | -2.55609 | 4.40847 | 2.67413 |
| C | 1.22205 | 1.75281 | 4.65986 | H | -2.52931 | -3.85239 | -1.74859 |
| C | 0.88255 | 1.08003 | 3.34078 | H | -3.06492 | -3.92396 | -0.27438 |
| O | 1.00444 | -0.18092 | 3.29127 | H | -3.85740 | -0.98218 | -2.71932 |
| O | 0.51860 | 1.85354 | 2.38930 | H | -2.96164 | 0.31028 | -2.79420 |
| O | 0.32023 | -2.21285 | 0.08310 | H | -2.41164 | 2.44020 | -3.33163 |
| Ca | -1.86527 | -1.52324 | -0.59413 | H | -4.20971 | 3.90353 | -0.30279 |
| Mn | 1.68933 | -1.36058 | -0.98925 | H | -3.01818 | 4.64679 | 0.38144 |
| O | -0.92898 | -0.25156 | 1.37679 | C | 1.16072 | 5.81891 | -0.68194 |
| Mn | 0.54659 | -1.28763 | 1.60484 | C | 1.29736 | 4.62647 | -1.64531 |
| O | 1.45995 | 0.03324 | 0.55275 | C | 0.42360 | 3.46429 | -1.18703 |
| Mn | -0.09785 | 1.17814 | 0.63292 | O | 0.93215 | 2.66163 | -0.33895 |
| O | -1.55789 | 2.32222 | 0.74678 | O | -0.75284 | 3.40689 | -1.67678 |
| Mn | -2.30877 | 2.31233 | -0.92165 | H | 1.46392 | 5.53655 | 0.33588 |
| H | 0.11979 | 6.16853 | -0.63941 | H | -2.21929 | -3.62904 | 4.32858 |
| H | 2.34343 | 4.28771 | -1.68385 | H | -0.64294 | -5.44168 | -5.44455 |
| H | 0.98335 | 4.91948 | -2.65769 | H | -6.06610 | 2.61854 | 1.59439 |
| H | 1.79489 | 6.65528 | -1.01295 | H | -7.60158 | 1.72043 | 1.51696 |
| H | 4.40801 | 1.74931 | -4.47807 | H | 5.76513 | 1.58248 | -2.28857 |
| H | 7.02045 | 1.25342 | 1.81469 | H | 0.67169 | 1.25855 | 5.47264 |
| H | 4.23205 | -3.61024 | 2.90891 | | | | |

$S_1$

| | | | | | | | |
|---|---|---|---|---|---|---|---|
| C | -6.18112 | 3.05918 | 1.13271 | O | -0.05697 | -2.19128 | -0.08713 |
| C | -5.73667 | 1.59731 | 1.12514 | Ca | -2.20790 | -1.04792 | -0.51628 |
| C | -4.44238 | 1.28914 | 0.35486 | Mn | 1.23817 | -1.37443 | -1.04678 |
| O | -4.04232 | 0.09500 | 0.38168 | O | -0.93599 | -0.22118 | 1.45582 |
| O | -3.87397 | 2.24735 | -0.26709 | Mn | 0.36142 | -1.47012 | 1.54633 |
| C | -1.64341 | -4.29609 | -4.70566 | O | 1.43916 | -0.29347 | 0.51587 |
| C | -0.31704 | -3.67090 | -4.27952 | Mn | 0.05344 | 1.17321 | 0.70221 |
| C | -0.40289 | -2.76692 | -3.04789 | O | -1.14432 | 2.50313 | 0.88641 |
| O | -1.48168 | -2.60361 | -2.45009 | Mn | -1.69844 | 2.50889 | -0.89583 |
| O | 0.72805 | -2.20713 | -2.73575 | O | -0.72532 | 0.87438 | -0.92411 |
| C | 3.56746 | 1.39556 | -3.31530 | O | -1.47505 | 4.89417 | 1.84838 |
| N | 4.55016 | 0.64441 | -2.69256 | O | -3.91527 | -2.92407 | -1.20210 |
| C | 2.37486 | 0.82643 | -2.94354 | O | -3.17882 | -0.10292 | -2.60135 |
| C | 3.94682 | -0.32606 | -1.96345 | O | -2.30267 | 2.37312 | -2.69603 |
| N | 2.62878 | -0.24062 | -2.10350 | O | -2.59457 | 4.48110 | -0.62293 |
| C | 6.29660 | 0.16348 | 1.14900 | H | -5.58333 | 1.22449 | 2.15116 |
| N | 6.04124 | 1.42121 | 0.63408 | H | -6.52026 | 0.95062 | 0.69564 |
| C | 5.11444 | -0.48019 | 1.46225 | H | -6.34878 | 3.43037 | 0.11143 |
| C | 4.71443 | 1.52595 | 0.64521 | H | -2.05074 | -4.94329 | -3.91580 |
| N | 4.11288 | 0.40793 | 1.14012 | H | -2.39503 | -3.52247 | -4.91549 |
| C | 3.61563 | -4.37243 | 1.15671 | H | 0.43932 | -4.44300 | -4.05990 |
| C | 2.69501 | -3.20717 | 0.81588 | H | 0.11509 | -3.06275 | -5.09135 |
| O | 1.91813 | -2.81406 | 1.74395 | H | 1.35929 | 1.11885 | -3.18669 |
| O | 2.76794 | -2.73976 | -0.36515 | H | 4.46914 | -1.04805 | -1.34390 |
| C | -2.48985 | -4.16555 | 3.22114 | H | 4.89870 | -1.45426 | 1.88737 |
| C | -1.90797 | -3.16321 | 2.21907 | H | 3.09335 | 0.19951 | 1.08702 |
| O | -2.56090 | -2.83465 | 1.20357 | H | 4.12824 | 2.37760 | 0.31226 |
| C | -3.79571 | -4.82307 | 2.77969 | H | 4.57796 | -4.26886 | 0.63836 |
| O | -0.74138 | -2.72425 | 2.55504 | H | 3.14039 | -5.29987 | 0.80042 |
| C | 1.65113 | 1.08355 | 4.75853 | H | -4.56453 | -4.06808 | 2.56535 |
| C | 1.13496 | 0.61674 | 3.40764 | H | -4.17736 | -5.49490 | 3.56438 |
| O | 1.03561 | -0.64728 | 3.25104 | H | -3.65396 | -5.41682 | 1.86468 |
| O | 0.88008 | 1.51301 | 2.54652 | H | -1.70893 | -4.91373 | 3.43083 |

| | | | | | | | | |
|---|---|---|---|---|---|---|---|---|
| H | 2.75188 | 1.07513 | 4.73452 | | O | 0.13631 | 3.63925 | -1.46321 |
| H | 1.31652 | 2.10786 | 4.96149 | | H | 2.80969 | 4.67064 | 1.07452 |
| H | -1.20174 | 3.93898 | 1.62878 | | H | 1.69459 | 5.88431 | 0.41447 |
| H | -2.08931 | 4.79494 | 2.59486 | | H | 3.27893 | 3.82685 | -1.26552 |
| H | -3.22029 | -3.19896 | -1.83977 | | H | 2.13475 | 5.02566 | -1.92573 |
| H | -3.66856 | -3.33691 | -0.34637 | | H | 3.43419 | 5.97797 | 0.03318 |
| H | -4.12758 | -0.01470 | -2.40740 | | H | 3.79876 | 2.24760 | -3.94314 |
| H | -2.85099 | 0.85766 | -2.72446 | | H | 7.31312 | -0.20114 | 1.27875 |
| H | -1.55077 | 2.52513 | -3.29510 | | H | 3.76161 | -4.45332 | 2.24076 |
| H | -3.46379 | 4.05746 | -0.42133 | | H | -2.62824 | -3.61312 | 4.16594 |
| H | -2.25287 | 4.80247 | 0.26293 | | H | -1.51133 | -4.90704 | -5.61226 |
| C | 2.59128 | 5.28784 | 0.19203 | | H | -5.41857 | 3.70354 | 1.59439 |
| C | 2.36717 | 4.40673 | -1.04738 | | H | -7.11737 | 3.18304 | 1.70060 |
| C | 1.20831 | 3.42857 | -0.83770 | | H | 5.54090 | 0.86020 | -2.65358 |
| O | 1.43916 | 2.47781 | -0.00366 | | H | 1.32188 | 0.39884 | 5.55020 |

$S_2$

| | | | | | | | | |
|---|---|---|---|---|---|---|---|---|
| C | 6.49804 | 2.35684 | -0.54105 | | C | 1.59042 | -3.37619 | -2.11746 |
| C | 5.80126 | 1.04873 | -0.91450 | | O | 2.24934 | -3.09562 | -1.09200 |
| C | 4.39099 | 0.85324 | -0.33568 | | C | 3.32618 | -5.21217 | -2.61677 |
| O | 3.91887 | -0.30571 | -0.33551 | | O | 0.48385 | -2.82434 | -2.49282 |
| O | 3.83436 | 1.92977 | 0.08364 | | C | -1.40767 | 1.16077 | -4.84472 |
| C | 1.03953 | -4.51143 | 4.70169 | | C | -1.06109 | 0.67417 | -3.44895 |
| C | -0.18246 | -3.68892 | 4.29942 | | O | -1.09844 | -0.59229 | -3.26471 |
| C | 0.03087 | -2.77981 | 3.08713 | | O | -0.75965 | 1.55594 | -2.59022 |
| O | 1.13715 | -2.71955 | 2.52086 | | O | -0.19506 | -2.17175 | 0.11742 |
| O | -1.02633 | -2.09954 | 2.75490 | | Ca | 2.02207 | -1.24259 | 0.59034 |
| C | -3.60550 | 1.53667 | 3.48411 | | Mn | -1.42918 | -1.23839 | 1.04597 |
| N | -4.56813 | 1.13142 | 2.57530 | | O | 0.91362 | -0.33540 | -1.50157 |
| C | -2.47908 | 0.81621 | 3.17388 | | Mn | -0.50782 | -1.44134 | -1.54456 |
| C | -4.01508 | 0.21039 | 1.74958 | | O | -1.43268 | -0.13972 | -0.51624 |
| N | -2.74841 | -0.00067 | 2.09139 | | Mn | 0.12735 | 1.16488 | -0.75962 |
| C | -6.13695 | 0.89318 | -1.36326 | | O | 1.35178 | 2.43808 | -1.02633 |
| N | -5.76301 | 2.12226 | -0.85192 | | Mn | 2.05990 | 2.55180 | 0.68577 |
| C | -5.02121 | 0.13275 | -1.65981 | | O | 0.97445 | 0.88916 | 0.80636 |
| C | -4.43222 | 2.09477 | -0.84831 | | O | 2.30877 | 4.43713 | -2.55585 |
| N | -3.93993 | 0.92005 | -1.33236 | | O | 3.56457 | -3.24571 | 1.34235 |
| C | -4.05241 | -3.98091 | -1.19488 | | O | 2.99637 | -0.29126 | 2.65942 |
| C | -3.02146 | -2.91775 | -0.83933 | | O | 2.66631 | 2.30651 | 2.47144 |
| O | -2.18592 | -2.61675 | -1.75099 | | O | 2.69826 | 4.24955 | 0.44324 |
| O | -3.07015 | -2.43392 | 0.33565 | | H | 5.68477 | 0.97002 | -2.01017 |
| C | 2.08203 | -4.45907 | -3.08272 | | H | 6.39386 | 0.17011 | -0.61576 |
| H | 6.66495 | 2.42145 | 0.54398 | | H | 2.87871 | 0.72676 | 2.63026 |
| H | 1.34632 | -5.18792 | 3.89117 | | H | 1.94473 | 2.62824 | 3.04236 |
| H | 1.89598 | -3.86165 | 4.92891 | | H | 2.56123 | 4.51153 | -0.49752 |
| H | -1.04575 | -4.33513 | 4.07006 | | C | -1.51408 | 5.75567 | 0.78904 |
| H | -0.51534 | -3.04136 | 5.12780 | | C | -1.74211 | 4.39435 | 1.46836 |
| H | -1.50100 | 0.82894 | 3.64134 | | C | -0.73999 | 3.36734 | 0.95032 |
| H | -4.54323 | -0.27293 | 0.93321 | | O | -1.12934 | 2.56953 | 0.03975 |
| H | -4.89719 | -0.85951 | -2.07974 | | O | 0.42363 | 3.44305 | 1.45688 |
| H | -2.95473 | 0.59941 | -1.23931 | | H | -1.66179 | 5.68185 | -0.29799 |
| H | -3.76874 | 2.88713 | -0.51437 | | H | -0.48852 | 6.10644 | 0.96947 |
| H | -5.00516 | -3.77938 | -0.68819 | | H | -2.76184 | 4.02968 | 1.28004 |
| H | -3.67911 | -4.95290 | -0.83595 | | H | -1.60017 | 4.49174 | 2.55534 |
| H | 4.16273 | -4.52212 | -2.44028 | | H | -2.21678 | 6.50675 | 1.18084 |
| H | 3.63876 | -5.94986 | -3.37223 | | H | -3.80248 | 2.28257 | 4.24479 |
| H | 3.13854 | -5.74907 | -1.67542 | | H | -7.18314 | 0.63123 | -1.50376 |
| H | 1.23832 | -5.14411 | -3.26433 | | H | -4.19166 | -4.04499 | -2.28093 |
| H | -1.97958 | 2.09551 | -4.78608 | | H | 2.26370 | -3.95653 | -4.04757 |
| H | -0.46753 | 1.37560 | -5.37590 | | H | 0.82182 | -5.12203 | 5.59170 |
| H | 1.80112 | 3.66976 | -2.16818 | | H | 5.88456 | 3.22386 | -0.81932 |
| H | 3.18797 | 4.05189 | -2.71112 | | H | 7.47509 | 2.43588 | -1.04441 |
| H | 2.82358 | -3.46849 | 1.94777 | | H | -5.49441 | 1.52738 | 2.45135 |
| H | 3.31914 | -3.63424 | 0.47459 | | H | -1.96303 | 0.39597 | -5.40025 |
| H | 3.95623 | -0.44117 | 2.66947 | | | | | |

$S_3$

| | | | | | | | | |
|---|---|---|---|---|---|---|---|---|
| C | 6.50874 | 1.94899 | -1.01160 | | C | -4.70899 | 0.70627 | -1.98858 |
| C | 5.82206 | 0.62622 | -0.67490 | | C | -4.26934 | 2.29734 | -0.53475 |
| C | 4.43811 | 0.70950 | -0.01710 | | N | -3.70914 | 1.26834 | -1.22839 |
| O | 3.88341 | -0.34735 | 0.31983 | | C | -3.95208 | -3.38952 | -2.08432 |
| O | 3.96645 | 1.90969 | 0.12772 | | C | -2.88197 | -2.48017 | -1.49516 |
| C | 0.08434 | -5.64532 | 3.79005 | | O | -1.93126 | -2.14852 | -2.27840 |
| C | -1.05538 | -4.73093 | 3.34490 | | O | -3.00876 | -2.13830 | -0.28429 |
| C | -0.62023 | -3.55734 | 2.45439 | | C | 2.51084 | -3.89426 | -3.24725 |
| O | 0.59298 | -3.32457 | 2.27043 | | C | 1.87645 | -2.92990 | -2.23569 |
| O | -1.62183 | -2.89565 | 1.97168 | | O | 2.38844 | -2.76545 | -1.11015 |
| C | -4.00857 | 1.11126 | 3.28003 | | C | 3.67489 | -4.71656 | -2.69848 |
| N | -4.93054 | 0.60033 | 2.37880 | | O | 0.81776 | -2.34362 | -2.70138 |
| C | -2.80864 | 0.52858 | 2.95314 | | C | -1.14895 | 1.91287 | -4.69539 |
| C | -4.27815 | -0.24349 | 1.53946 | | C | -0.75320 | 1.23839 | -3.39324 |
| N | -2.99492 | -0.30633 | 1.87023 | | O | -0.71892 | -0.02891 | -3.37510 |
| C | -5.84557 | 1.44501 | -1.71649 | | O | -0.50119 | 2.01710 | -2.41491 |
| N | -5.56613 | 2.44191 | -0.80052 | | O | -0.17112 | -2.11084 | -0.19689 |

| | | | | | | | |
|---|---|---|---|---|---|---|---|
| Ca | 1.71793 | -1.22931 | 1.00243 | H | 1.70696 | -4.54050 | -3.63445 |
| Mn | -1.42308 | -1.28521 | 0.87449 | H | -2.09950 | 2.44642 | -4.54817 |
| O | 1.13071 | -0.01993 | -1.37571 | H | -0.38912 | 2.65913 | -4.96648 |
| Mn | -0.30023 | -1.10394 | -1.68974 | H | 2.37658 | -3.71594 | 1.84243 |
| O | -1.23717 | 0.12915 | -0.55538 | H | 3.15576 | -3.45475 | 0.51289 |
| Mn | 0.29706 | 1.43201 | -0.66793 | H | 3.76567 | -0.56627 | 2.92269 |
| O | 1.56864 | 2.72857 | -0.88139 | H | 2.88853 | 0.72079 | 2.92267 |
| Mn | 2.25194 | 2.59419 | 0.81073 | H | 2.36115 | 2.88965 | 3.16386 |
| O | 1.22297 | 1.16701 | 0.91701 | H | 2.76420 | 4.37208 | -0.67711 |
| O | 3.18882 | -3.29047 | 1.48292 | C | -1.76371 | 5.76068 | 0.73024 |
| O | 2.85132 | -0.27134 | 3.07869 | C | -1.35181 | 4.68955 | 1.75285 |
| O | 2.99458 | 2.46920 | 2.55281 | C | -0.39720 | 3.67071 | 1.13963 |
| O | 3.30650 | 4.45212 | 0.14258 | O | -0.87460 | 2.85822 | 0.30045 |
| H | 5.68264 | 0.01430 | -1.58111 | O | 0.82648 | 3.75528 | 1.52665 |
| H | 6.44604 | 0.01599 | -0.00205 | H | -2.23348 | 5.30042 | -0.14982 |
| H | 6.66439 | 2.56003 | -0.11044 | H | -0.88944 | 6.33497 | 0.38964 |
| H | 0.57492 | -6.11811 | 2.92671 | H | -2.24094 | 4.14209 | 2.10302 |
| H | 0.85393 | -5.07967 | 4.33321 | H | -0.86296 | 5.14995 | 2.62266 |
| H | -1.83227 | -5.29068 | 2.80017 | H | -2.48137 | 6.46519 | 1.17672 |
| H | -1.56637 | -4.28916 | 4.21738 | H | -4.28833 | 1.81834 | 4.05239 |
| H | -1.81550 | 0.63053 | 3.37922 | H | -6.84162 | 1.32207 | -2.13615 |
| H | -4.73767 | -0.77046 | 0.71079 | H | -3.92794 | -3.36946 | -3.18053 |
| H | -4.52036 | -0.13652 | -2.64505 | H | 2.83275 | -3.28016 | -4.10540 |
| H | -2.74358 | 0.88599 | -1.07975 | H | -0.29072 | -6.44416 | 4.44943 |
| H | -3.68147 | 2.90454 | 0.14761 | H | 5.90942 | 2.53932 | -1.71969 |
| H | -4.94463 | -3.10215 | -1.71225 | H | 7.49301 | 1.76897 | -1.47224 |
| H | -3.75361 | -4.41817 | -1.74543 | H | -5.88651 | 0.91529 | 2.25154 |
| H | 4.46824 | -4.06546 | -2.30588 | H | -1.26047 | 1.17360 | -5.49665 |
| H | 4.10652 | -5.35481 | -3.48570 | O | -0.33001 | -0.58965 | 1.96240 |
| H | 3.34900 | -5.36757 | -1.87410 | H | 4.10933 | 3.92318 | -0.06182 |

$S_3OOH$

| | | | | | | | |
|---|---|---|---|---|---|---|---|
| C | -7.15766 | -1.34529 | 0.19618 | C | 3.50208 | 2.78702 | -2.93271 |
| C | -5.87834 | -2.10165 | -0.16308 | N | 4.31668 | 2.80403 | -1.80987 |
| C | -4.68135 | -1.23953 | -0.57828 | C | 2.70174 | 1.67971 | -2.79499 |
| O | -3.56491 | -1.79677 | -0.68634 | C | 3.98072 | 1.74504 | -1.02966 |
| O | -4.90990 | 0.00840 | -0.81510 | N | 3.00696 | 1.04820 | -1.60232 |
| C | 5.02593 | -4.77601 | -2.91368 | C | 3.98731 | 4.26237 | 2.03389 |
| C | 4.86039 | -3.31694 | -2.49008 | N | 3.27361 | 4.96404 | 1.07878 |
| C | 3.48051 | -2.98121 | -1.87650 | C | 3.43827 | 3.00720 | 2.22264 |
| O | 2.58668 | -3.85045 | -1.90860 | C | 2.30469 | 4.12609 | 0.70702 |
| O | 3.43298 | -1.79055 | -1.39341 | N | 2.36280 | 2.93897 | 1.36954 |

| | | | | | | | | |
|---|---|---|---|---|---|---|---|---|
| C  | 4.70530  | -1.09047 | 3.00926  | | H  | 1.52749  | 4.31968  | -0.02734 |
| C  | 3.50585  | -0.72423 | 2.13027  | | H  | 5.55025  | -0.41776 | 2.80688  |
| O  | 2.36104  | -0.92661 | 2.65821  | | H  | 5.01902  | -2.11370 | 2.74871  |
| O  | 3.73371  | -0.28174 | 0.97485  | | H  | -2.50789 | -5.48058 | 2.27951  |
| C  | -1.08405 | -4.32827 | 3.45211  | | H  | -1.93910 | -6.31192 | 3.75797  |
| C  | -0.71533 | -3.31228 | 2.35542  | | H  | -0.88696 | -6.18528 | 2.32358  |
| O  | -1.13034 | -3.49760 | 1.18658  | | H  | -0.19388 | -4.47979 | 4.08260  |
| C  | -1.63723 | -5.65252 | 2.92702  | | H  | -0.38586 | 3.36324  | 4.37434  |
| O  | -0.00516 | -2.33405 | 2.79179  | | H  | -1.75432 | 2.29427  | 4.74277  |
| C  | -0.67757 | 2.31338  | 4.51569  | | H  | 0.91136  | -3.92363 | -1.54370 |
| C  | -0.44031 | 1.53633  | 3.22785  | | H  | -0.04817 | -4.30784 | -0.36648 |
| O  | 0.15830  | 0.42184  | 3.32702  | | H  | -0.36489 | 0.38988  | -2.54437 |
| O  | -0.87082 | 2.07614  | 2.16124  | | H  | -2.91768 | 1.08105  | -3.68960 |
| O  | 1.13217  | -1.80795 | 0.32799  | | H  | -4.40789 | 1.78949  | 1.42357  |
| Ca | -1.12826 | -1.94692 | -0.85219 | | C  | -1.69436 | 5.91840  | -0.57252 |
| Mn | 2.00587  | -0.57065 | -0.65533 | | C  | -1.24971 | 4.96620  | -1.69391 |
| O  | -1.14795 | -0.54845 | 1.14014  | | C  | -1.55223 | 3.50691  | -1.32984 |
| Mn | 0.56019  | -0.76185 | 1.71387  | | O  | -0.75899 | 2.98653  | -0.47274 |
| O  | 0.97434  | 0.67032  | 0.53384  | | O  | -2.55892 | 2.97324  | -1.88366 |
| Mn | -1.05426 | 1.12885  | 0.37315  | | H  | -1.18384 | 5.66995  | 0.36843  |
| O  | -2.83984 | 1.41484  | 0.41757  | | H  | -2.77819 | 5.83747  | -0.40021 |
| Mn | -3.16453 | 0.94574  | -1.32653 | | H  | -0.16541 | 5.06706  | -1.86101 |
| O  | -1.32429 | 0.45341  | -1.28949 | | H  | -1.77096 | 5.20442  | -2.63286 |
| O  | -0.04984 | -4.11530 | -1.32599 | | H  | -1.46318 | 6.96537  | -0.82722 |
| O  | 0.30873  | -0.05547 | -3.16627 | | H  | 3.56325  | 3.54780  | -3.70241 |
| O  | -3.45999 | 0.47593  | -3.15146 | | H  | 4.84492  | 4.70234  | 2.53976  |
| O  | -5.39021 | 1.80267  | 1.54254  | | H  | 4.43775  | -1.07011 | 4.07383  |
| H  | -5.53652 | -2.73014 | 0.67412  | | H  | -1.82641 | -3.82671 | 4.09713  |
| H  | -6.05734 | -2.79569 | -1.00272 | | H  | 6.01383  | -4.94799 | -3.37490 |
| H  | -7.48967 | -0.71339 | -0.63971 | | H  | -6.99822 | -0.68883 | 1.06424  |
| H  | 4.92537  | -5.45078 | -2.05057 | | H  | -7.97064 | -2.04785 | 0.44537  |
| H  | 4.24650  | -5.06299 | -3.63344 | | H  | 4.91842  | 3.56435  | -1.51312 |
| H  | 5.63017  | -3.01971 | -1.75946 | | H  | -0.11927 | 1.86829  | 5.34813  |
| H  | 4.99351  | -2.64209 | -3.35317 | | O  | 0.79347  | -1.06682 | -2.19890 |
| H  | 1.90890  | 1.27822  | -3.42301 | | H  | -5.64688 | 1.24729  | 0.77968  |
| H  | 4.43359  | 1.50439  | -0.07365 | | O  | -1.91210 | -1.84125 | -3.26395 |
| H  | 3.70554  | 2.17651  | 2.86640  | | H  | -2.57066 | -1.08431 | -3.26521 |
| H  | 1.78637  | 2.08315  | 1.14888  | | H  | -1.08221 | -1.39805 | -3.55953 |

$S_4OOH + H_2O$

| | | | | | | | |
|---|---|---|---|---|---|---|---|
| C | 7.28709 | 0.65419 | 0.42507 | C | 4.81611 | 0.95752 | -0.31250 |
| C | 6.13942 | 1.61051 | 0.10388 | O | 3.82021 | 1.71144 | -0.44536 |
| O  |  4.82054 | -0.30713 | -0.50997 | H  | -3.77049 |  5.81464 | -2.84348 |
| C  | -4.00800 |  5.03726 | -3.58358 | H  | -3.19757 |  5.04113 | -4.32538 |
| C  | -4.13603 |  3.66723 | -2.92090 | H  | -4.95536 |  3.65014 | -2.18414 |
| C  | -2.86088 |  3.19336 | -2.20196 | H  | -4.39179 |  2.89097 | -3.66174 |
| O  | -1.83178 |  3.88407 | -2.25374 | H  | -2.31163 | -0.84692 | -3.58610 |
| O  | -3.03890 |  2.06140 | -1.59172 | H  | -4.39264 | -1.22218 |  0.05430 |
| C  | -4.02700 | -2.16265 | -3.04563 | H  | -3.74465 | -1.55252 |  3.09908 |
| N  | -4.67240 | -2.24540 | -1.82319 | H  | -1.95370 | -1.92965 |  1.29893 |
| C  | -3.05620 | -1.20581 | -2.88698 | H  | -2.29463 | -4.14991 |  0.11215 |
| C  | -4.08401 | -1.36908 | -0.97391 | H  | -5.49488 |  0.92913 |  2.29957 |
| N  | -3.09796 | -0.72470 | -1.58956 | H  | -4.92949 |  2.61189 |  2.32283 |
| C  | -4.59802 | -3.46451 |  2.23830 | H  |  3.13922 |  5.15337 |  2.38043 |
| N  | -4.11637 | -4.31463 |  1.26054 | H  |  2.49609 |  6.20720 |  3.67130 |
| C  | -3.73105 | -2.40457 |  2.42784 | H  |  1.67232 |  6.09948 |  2.09542 |
| C  | -2.96846 | -3.76469 |  0.87181 | H  |  0.44181 |  4.71985 |  3.84721 |
| N  | -2.69132 | -2.61696 |  1.55065 | H  | -0.19739 | -3.24635 |  4.52295 |
| C  | -4.67105 |  1.58246 |  2.61532 | H  |  1.24436 | -2.30214 |  4.94648 |
| C  | -3.38377 |  1.19859 |  1.90459 | H  | -0.22844 |  3.87461 | -1.76201 |
| O  | -2.30939 |  1.27015 |  2.56839 | H  |  0.71227 |  4.32317 | -0.58834 |
| O  | -3.49177 |  0.84863 |  0.68397 | H  |  0.42019 | -0.60608 | -2.21514 |
| C  |  1.37656 |  4.36449 |  3.38453 | H  |  3.07412 | -1.73445 | -3.48329 |
| C  |  0.99996 |  3.32074 |  2.32791 | H  |  4.37929 | -3.08946 |  0.31306 |
| O  |  1.47164 |  3.39288 |  1.17415 | C  |  0.75728 | -6.09721 | -0.62586 |
| C  |  2.21969 |  5.52218 |  2.85450 | C  |  0.33245 | -4.98068 | -1.59407 |
| O  |  0.18995 |  2.41357 |  2.77436 | C  |  0.96103 | -3.64508 | -1.19495 |
| C  |  0.18750 | -2.22569 |  4.64920 | O  |  0.30257 | -2.96624 | -0.32194 |
| C  |  0.11464 | -1.48346 |  3.32708 | O  |  2.06687 | -3.34139 | -1.71840 |
| O  | -0.34246 | -0.29372 |  3.36031 | H  |  0.44238 | -5.86454 |  0.40146 |
| O  |  0.51307 | -2.10704 |  2.29777 | H  |  1.85040 | -6.21859 | -0.63039 |
| O  | -0.94057 |  1.95787 |  0.31602 | H  | -0.76296 | -4.87646 | -1.59432 |
| Ca |  1.52820 |  1.78319 | -0.75898 | H  |  0.66029 | -5.21987 | -2.61614 |
| Mn | -1.90662 |  0.78353 | -0.64117 | H  |  0.30453 | -7.05794 | -0.91492 |
| O  |  1.19457 |  0.42581 |  1.26685 | H  | -4.30832 | -2.77510 | -3.89420 |
| Mn | -0.46609 |  0.90025 |  1.74371 | H  | -5.52925 | -3.66994 |  2.76123 |
| O  | -1.03884 | -0.48121 |  0.53834 | H  | -4.54030 |  1.54721 |  3.70333 |
| Mn |  0.94493 | -1.24964 |  0.50337 | H  |  1.90981 |  3.82026 |  4.18243 |
| O  |  2.63572 | -1.82260 |  0.63036 | H  | -4.94647 |  5.31522 | -4.08927 |
| Mn |  3.13786 | -1.49554 | -1.11351 | H  |  7.02647 | -0.01127 |  1.26118 |
| O  |  1.44252 | -0.61501 | -1.12478 | H  |  8.19458 |  1.21233 |  0.70576 |
| O  |  0.73434 |  3.94544 | -1.48902 | H  | -5.37190 | -2.92656 | -1.54740 |
| O  | -0.34347 | -0.21995 | -2.80687 | H  | -0.37125 | -1.69526 |  5.42861 |
| O  |  3.53826 | -1.08967 | -2.91841 | O  | -0.62689 |  0.94113 | -2.00868 |
| O  |  4.87516 | -2.93697 | -0.51927 | H  |  5.38707 | -2.11357 | -0.33783 |
| H  |  5.91194 |  2.26419 |  0.96107 | O  |  2.23789 |  1.34387 | -3.14401 |
| H  |  6.41655 |  2.29155 | -0.71865 | H  |  2.78148 |  0.49562 | -3.12125 |
| H  |  7.52910 |  0.02006 | -0.43952 | H  |  1.39197 |  1.05601 | -3.53977 |